\documentclass[12pt]{iopart}%
\usepackage{graphicx}
\usepackage{amssymb}
\usepackage{natbib}

\def\araa{{\em Ann.\ Rev.\ Astron.\ Astrophys.\ }}

\def\aaps{{\em Astron.\ Astrophys.\ Suppl.\ Ser.\ }}

\def\apj{{\em Astrophys.\ J.\ }}
\def\apjl{{\em Astrophys.\ J.\, Lett.\ }}

\def\mnras{{\em Mon.\ Not.\ R.\ Astron.\ Soc.\ }}
\def\nat{{\em Nature }}

\begin{document}

\title[The Inhomogeneous ${\rm H}_{2}$ Dissociating Background]{The
  Inhomogeneous Background of ${\rm H}_{2}$ Dissociating Radiation 
during Cosmic Reionization}

\author{Kyungjin Ahn$^1$, Paul R. Shapiro$^{2,\,3}$, Ilian T. Iliev$^{4,\,5}$,
  Garrelt Mellema$^6$ and Ue-Li Pen$^7$} 

\address{$^1$ Department of Earth Science Education, Chosun University, Gwangju
  501-759, Korea} 
\ead{$^1$ kjahn@chosun.ac.kr}

\address{$^2$ Department of Astronomy, University of Texas, Austin, TX 78712-1083, 
U.S.A}

\address{$^3$ Texas Cosmology Center, the University of Texas at
  Austin, TX 78712, U.S.A}

\address{$^4$ Universit\"at Z\"urich, Institut f\"ur Theoretische Physik,
Winterthurerstrasse 190, CH-8057 Z\"urich, Switzerland}

\address{$^5$ Current address: Astronomy Centre, Department of Physics
  \& Astronomy, University of Sussex, BRIGHTON BN1 9QH, England}

\address{$^6$ Dept. of Astronomy, AlbaNova University Center, Stockholm University,
SE 10691 Stockholm, Sweden}

\address{$^7$ Canadian Institute for Theoretical Astrophysics, University
 of Toronto, 60 St. George Street, Toronto, ON M5S 3H8, Canada}


\begin{abstract}
The first, self-consistent calculations are presented of the cosmological,
${\rm H}_2$-dissociating UV background produced during the epoch of 
reionization by the sources of reionization. Large-scale radiative 
transfer simulations of reionization trace the impact of all the 
ionizing starlight on the IGM from all the sources in our simulation 
volume down to dwarf galaxies of mass $\sim 10^8 \,{\rm M}_\odot$, 
identified by very high-resolution N-body simulations, including the 
self-regulating effect of IGM photoheating on dwarf galaxy formation. 
The UV continuum emitted below 13.6 eV by each source is then 
transferred through the same IGM, attenuated by atomic H Lyman series 
resonance lines, to predict the evolution of the inhomogeneous radiation
background in the Lyman-Werner bands of ${\rm H}_2$ between 11 and 
13.6~eV. On average, the intensity of this Lyman-Werner background is 
found to rise to the threshold level at which dissociation suppresses 
${\rm H_2}$ 
cooling and star formation inside minihalos, long before reionization 
is complete. Spatial variations in the Lyman-Werner background are 
found which result from the clustering of sources associated with 
large-scale structure formation, such that intensity fluctuations  
correlate with matter density fluctuations. As a result, the Lyman-Werner
background rises to the threshold level for ${\rm H_2}$ suppression earlier
in the vicinity of the reionization sources and their H~II regions.
\end{abstract}


\section{Introduction}

Simulations suggest that the first stars in the Cold Dark Matter
(``CDM'') universe formed inside minihalos of mass $M \sim 10^{5-6}
{\rm M}_\odot$ at $z\gtrsim 20$, when ${\rm H}_2$ molecules cooled the
primordial, metal-free halo gas and gravitational collapse ensued (e.g.
\citealt{2004ARA&A..42...79B}
and references therein). This critical role of ${\rm H}_2$ molecules
as the primary coolants responsible for triggering the gravitational
collapse that caused stars to form inside minihalos was limited,
however, by the fact that ${\rm H}_2$ can be dissociated by absorbing
UV radiation in the ${\rm H}_2$ Lyman-Werner (``LW'') bands in the
energy range $11-13.6\,$eV. In the presence of a high enough LW-band
radiation intensity, $J_{\rm LW}$, the ${\rm H}_2$ abundance would
have been too low to cool the gas sufficiently to form stars.
The threshold level of the intensity, $(J_{{\rm LW}})_{{\rm threshold}}$,
above which minihalo star formation was suppressed is still
uncertain. Early estimates (\citealt{2000ApJ...534...11H}; henceforth,
``HAR'') found that $(J_{{\rm LW}})_{{\rm threshold}}$ depended upon minihalo 
mass and redshift. In terms of the dimensionless quantity,
$J_{{\rm LW},\,21}\equiv J_{\rm LW}/(10^{-21}\,{\rm erg}\,{\rm
 s}^{-1}\,{\rm cm}^{-2}\,{\rm Hz^{-1}}\,{\rm sr}^{-1})$, they found that, to
suppress ${\rm H}_2$ for {\em all} minihalo masses, $\left( J_{{\rm
 LW},\,21} \right)_{\rm threshold} \sim 10^{-2}$ to $1$ was required,
as redshift varied from $z\sim 10$ to $50$, respectively. Later
estimates, including those based upon 3D, numerical gas dynamical
simulation of minihalos evolving in the presence of a LW background
from more realistic cosmological 
initial conditions, found a similar range of threshold values
(e.g. \citealt{2002ApJ...575...49R}; \citealt{2003ApJ...592..645Y}; 
\citealt{2007ApJ...663..687Y}).

Once stars began to form inside minihalos and also inside the rarer,
more massive halos with $M\gtrsim 10^8\,{\rm M}_\odot$ virial temperatures 
above $10^4 {\rm K}$, in which radiative cooling by atomic hydrogen 
was possible even without ${\rm H_2}$ (to cool the gas down to 
$10^4 {\rm K}$, at least, and initiate gravitational collapse), a rising
diffuse LW background would have been inevitable, however. Starlight at
energies below $13.6\,{\rm eV}$, the ionization potential of atomic
hydrogen, would have been largely free to escape from these star
forming halos into the intergalactic medium (``IGM''), while the
ionizing radiation above $13.6\,{\rm eV}$ was partially absorbed by
the neutral hydrogen within the halos and thereby reduced by an escape
fraction, $f_{\rm esc}$. In that case, given the value of this $f_{\rm
  esc}$ and the ratio of the number of ionizing photons to the number
of ${\rm H}_2$-dissociating photons released by the stars, $N_{\rm
  i}/N_{\rm LW}$,  which depends on the mass function and spectra of
the stars, the rise of the cosmic LW background can, in principle, be
related to the rise of the diffuse ionizing background. The latter is
believed to have been responsible for the reionization of the intergalactic
medium completed by redshift $z\gtrsim 6$. Estimates by HAR showed
that the sources of this reionization would have caused the mean LW
intensity in the IGM to exceed $\left( J_{{\rm LW},\,21} \right)_{\rm
  threshold}$ long before reionization was complete. 

This outcome is expected on quite general grounds, in fact.
The mean number of LW photons in the background per H atom is 
related to the LW intensity according to:
\begin{eqnarray}
\left( \frac{n_{\rm LW}}{n_{\rm H}}\right)
&=&\left( \frac{4\pi}{c}\int_{11.5\,{\rm eV}}^{13.6\,{\rm eV}}
\frac{J_{\nu}}{h\nu}d\nu \right)/n_{\rm H}, \nonumber \\ 
&\simeq &\frac{4\pi}{ch}\frac{2.1\,{\rm
    eV}}{12.6\,{\rm eV}}\langle
J_{\nu}\rangle /n_{\rm H},
\nonumber \\ 
&\simeq & 1.05\times 10^{-5} J_{{\rm LW},\,21}/n_{\rm H},  \nonumber \\ 
&\simeq & 6.3\times 10^{-3} J_{{\rm LW},\,21}
\left( \frac{1+z}{21} \right)^{-3}.
\label{eq:nLW_nH}
\end{eqnarray}
Hence, if $\left( J_{{\rm LW},\,21} \right)_{\rm threshold} <1$, this
implies that $(n_{\rm LW}/n_{\rm H})_{\rm threshold} \lesssim 6\times
10^{-3} (1+z)_{21}^{-3}\ll 1$.
The mean number of photons in the LW background per H atom at a given
epoch is determined by their rate of emission per atom, integrated
over time, reduced by a factor which accounts for attenuation and
redshifting after emission. As discussed in \S~\ref{sec:Methodology}, 
photons 
emitted in the LW bands below 13.6 eV are removed from the background
when they redshift to the frequency of the nearest H atom Lyman series
transition from level $n\ge 3$ to  $n=1$. 
Only those sources within the distance from which photons emitted at
the Ly$\gamma$ frequency would be received at the Ly$\beta$ frequency
can contribute at all to the LW background at a given point.
On average, roughly a
third of the LW photons emitted within this horizon will
survive their trip\footnote{This factor 1/3 comes
from the survival rate from the attenuation by Lyman series resonace lines,
given approximately by $\int_{0}^{r_{\rm LW}} dr_{\rm os} f_{\rm mod}
(r_{\rm os}) / \int_{0}^{r_{\rm LW}} dr_{\rm os}$, 
where the terms used are described in
\S~\ref{sub:picket-fence}. Compare this expression to equation
(\ref{eq:Jnuhom}) and also see Figure~\ref{fig:J21comp}.}.
The same sources also emit UV photons with energies above 13.6 eV
which are destroyed en-route by photoionizing H atoms. Reionization
required that at least one ionizing photon was released into the IGM
per H atom by the end of reionization. Let $\xi_{\rm LW}$ and
$\xi_{\rm i}$ be the total number of LW and ionizing photons, respectively, 
released per H atom into the IGM up to some time. The mean number of 
photons per H atom in the LW
background at that time is then roughly given by
\begin{equation}
\frac{n_{\rm LW}}{n_{\rm H}}\simeq \frac{1}{3}\left(\frac{\xi_{\rm
      LW}}{\xi_{\rm i}}\right) \xi_{\rm i} .
\label{eq:nn_xixi}
\end{equation}
For the stars believed to
be responsible for reionization, the intrinsic ratio of ionizing UV
photons to LW photons released, $N_{\rm i}/N_{\rm LW}$, ranged from
$N_{\rm i}/N_{\rm LW} \sim 15$ (Pop III stars, high mass) to 
$N_{\rm i}/N_{\rm LW} \sim 1$ (Pop II stars, Salpeter IMF), while
the escape fraction of ionizing photons was some
$f_{\rm esc} \ll 1$. 
In that case, we can write
\begin{equation}
\frac{n_{\rm LW}}{n_{\rm H}}\simeq \frac{1}{3} 
f_{\rm esc}^{-1}\left(\frac{N_{\rm 
      i}}{N_{\rm LW}}\right)^{-1} \xi_{\rm i},
\label{eq:nn_NNxi}
\end{equation}
and, therefore, if ($J_{\rm LW,\,21})_{\rm threshold}<1$, 
equations~(\ref{eq:nLW_nH}) and (\ref{eq:nn_NNxi}) imply that
$\left( n_{\rm LW}/n_{\rm H}\right)$ reaches the threshold level 
when $\xi_{\rm i}$ is only
\begin{equation}
\left( \xi_{\rm i} \right)_{\rm LW,\, threshold}
\simeq 3 \left( \frac{n_{\rm LW}}{n_{\rm H}} \right)_{\rm threshold} 
\left(\frac{N_{\rm i}}{N_{\rm LW}}\right)
f_{\rm esc} \ll 1 .
\label{eq:xi_th}
\end{equation}
Accordingly, since reionization was not complete until the condition
$\xi_{\rm i}>1$ was reached, the LW threshold for
suppressing ${\rm H}_2$ in the minihalos must have been
reached long before the end of reionization. 
As a result, as HAR suggested, the minihalos were generally
``sterilized'' before they could contribute significantly to reionization.

What level of LW background is required to suppress
minihalo star formation when the minihalos are also directly exposed
to {\it ionizing} radiation, as well, is a more complicated question to
answer. If a minihalo forms in a region of the IGM which is already
photoionized, the pressure of the photoheated IGM there prevents it 
from collapsing gravitationally into the dark-matter dominated halo. 
Those minihalos are missing their baryonic component, therefore.
This phenomenon, sometimes referred to as ``Jeans-mass filtering''
\citep{1994ApJ...427...25S,1998MNRAS.296...44G}, ensures
that ${\rm H}_2$ cooling and star formation do not occur inside
minihalos in the ionized regions of the IGM unless those minihalos
had formed there {\it prior} to the arrival of the ionizing radiation. 
The impact of both LW {\it and} ionizing radiation (including X-rays)
on {\it pre-existing}
minihalos inside H~II regions is a subject of ongoing work, beyond the
scope of this paper \citep[e.g.][]{2001ApJ...548..509M,2002ApJ...575...49R,
2003ApJ...592..645Y,2003MNRAS.346..4560,2004MNRAS.348..753S,
2005MNRAS...361..405I,2006ApJ...639..621A,2006ApJ...650....7H,
2006ApJ...645L..93S,2006ApJ...648..835M,2007MNRAS.375..881A,
2007ApJ...671.1559W,2007ApJ...663..687Y,2008ApJ...673...14O}. The presence
of dissociating radiation {\it always} implies the potential for
limiting the ${\rm H_2}$ abundance and, with it, the cooling required 
to make stars form. This is true even for the atomic-cooling halos above
the minihalo mass range, although the level required for suppression might
be higher. We shall focus here, however, on the rise of the LW background 
and its spatial variations contributed by the dominant sources of 
reionization, but leave the question of how the intensity impacts star
formation for future studies.

Previous estimates of the cosmic LW background were limited to the 
{\it mean} background and were based upon a {\it homogeneous} 
approximation. These calculations assumed that the sources and the IGM
were uniformly distributed, with uniform emissivity, given either by 
analytical approximation (HAR) or by summing over the sources found in 
small-box simulations, too small to account for the large-scale clustering 
of sources or to follow global reionization \citep[e.g.][]{2002ApJ...575...49R,
2003ApJ...592..645Y}.
Background of Ly$\alpha$ pumping radiation was considered semi-analytically 
by \citet{2005ApJ...626....1B} and \citet{2006MNRAS.367.1057P}, including 
the effects of fluctuations in the background in a linear approximation.
The focus of these papers is on the Ly$\alpha$ intensity originating
from photons that are absorbed by hydrogen Lyman resonance lines and
converted into Ly$\alpha$ photons, which then radiatively mixes the 21
cm levels to drive the spin temperature to the gas kinetic temperature
(Wouthuysen-Field effect), while we are interested in photons that
remain unattenuated by Lyman resonance lines and affect ${\rm H}_2$
abundance through photo-dissociation. Note that the horizon for sources
responsible for the fluctuating Ly$\alpha$ background is considerably
larger than that responsible for the LW background.

Here we present the first self-consistent radiative transfer calculations
of the {\it inhomogeneous} LW background produced by the same sources which 
reionized the universe in a large-scale radiative transfer simulation of 
reionization. This problem presents a formidable computational challenge.
The horizon for seeing LW photons is $\sim100$ comoving Mpc, much larger 
than the size of typical H~II regions ($\sim10$ Mpc). Since the mean free 
path for LW photons ($\sim100$ Mpc) is much larger than that for H-ionizing
photons we must account for sources distributed over large volume and 
lookback time. Finally, the LW band photons are attenuated as they redshift
into H-atom Lyman series resonance lines as they travel across the IGM. Hence,
it is necessary to perform a multi-frequency radiative transfer calculation 
from each of the millions of sources in a cosmological volume larger than 
$\sim (100\, \rm Mpc)^3$, integrated along the light cones from each source
to every observation point they intersect, which is computationally 
prohibitive. As we shall show, a good approximation is possible which 
reduces the multi-frequency calculation to an equivalent gray opacity 
calculation. As a result, we are not only able to derive the evolution of 
the rising globally-averaged mean LW background during the epoch of 
reionization (EOR), but also to map 
its pattern of spatial variations over time. In \S~\ref{sec:Methodology}, 
we describe how continuum radiation emitted in the LW range below 13.6~eV 
is transferred through the IGM along the light cones from sources to 
observers, and how we solve
this problem numerically. In \S~\ref{sec:applic}, we apply this method to 
one of our recent large-scale radiative transfer simulations of 
self-regulated reionization, described in \citet{2007MNRAS.376..534I}. 
We compare the mean LW background evolution thus derived numerically with 
the homogeneous universe approximation and describe the spatial 
fluctuations of the LW background in some detail. Our conclusions are 
summarized in \S~\ref{sec:conclusions}. Some of our results on the 
inhomogeneous LW background during the EOR described here were first 
summarized in \citet{2008AIPC..990..374A}.

\section{Radiative Transfer of the LW Background}
\label{sec:Methodology}

\subsection{Basic Equations}
\label{sub:Basics}
Let us first briefly describe how the inhomogeneous LW background can
be calculated. Consider radiation sources distributed inhomogeneously.
The mean intensity $J_\nu({\bf x}_{\rm obs},z_{\rm obs},\nu_{\rm obs})$ at 
observed frequency $\nu_{\rm obs}$ at some comoving position 
${\bf x}_{\rm obs}$ at redshift $z_{\rm obs}$ is given by
\begin{equation}
J_\nu({\bf x}_{\rm obs},z_{\rm obs},\nu_{\rm obs}) = 
\frac{1}{4\pi}\sum_{{\rm s}} 
          F_{\nu,\,{\rm s}}({\bf x}_{\rm obs},z_{\rm obs},\nu_{\rm obs}),
\label{eq:bruteforce}
\end{equation}
where $F_{\nu,\,{\rm s}}$ is the flux received at 
$({\bf x}_{\rm obs},z_{\rm obs},\nu_{\rm obs})$ that was emitted at 
$({\bf x}_{\rm s},z_{\rm s},\nu_{\rm s})$ by a source (denoted by 
subscript $s$), where 
\begin{equation}
\frac{\nu_{\rm s}}{\nu_{\rm obs}}=\frac{1+z_{\rm s}}{1+z_{\rm obs}}.
\label{equ:redshifting}
\end{equation}
The position and redshift, $({\bf x}_{\rm s},z_{\rm s})$, of a source 
are related to those of the observer at $({\bf x}_{\rm obs},z_{\rm obs})$
by the fact that the signal emitted at the epoch $z_{\rm s}$ must reach 
the position ${\bf x}_{\rm obs}$ at the epoch $z_{\rm obs}$, travelling 
at the speed of light while the universe expands. We express this 
implicitly by writing the comoving separation, $r_{\rm os}$, of the source 
and observer as follows: 
\begin{equation}
r_{\rm os}=|{\bf x}_{\rm obs}-{\bf x}_{\rm s}|
 =\int_{t(z_{\rm s})}^{t(z_{\rm obs})}\frac{cdt}{a(t)}
 =-\int_{z_{\rm obs}}^{z_{\rm s}}c\frac{dz}{H(z)}.
\label{equ:r_os}
\end{equation}
If we specialize to the case of interest here ($z>6$) in which the Hubble
parameter $H(z)$ is given by the high-redshift limit for a flat universe 
with cosmological constant, equation~(\ref{equ:r_os}) can be integrated to
yield
\begin{eqnarray}
r_{\rm os}&=&
2cH_0^{-1}\Omega_m^{-1/2}[(1+z_{\rm obs})^{-1/2}-(1+z_{\rm s})^{-1/2}],\nonumber\\
&=&2cH_{0}^{-1}\Omega_{m}^{-1/2}(1+z_{{\rm obs}})^{-1/2}\left[\left(\frac{\nu_{{\rm obs}}}{\nu_{{\rm s}}}\right)^{-1/2}-1\right],
\label{eq:r_os2}
\end{eqnarray}
using equation~(\ref{equ:redshifting}).

The differential flux, $F_{\nu,s}$, received at 
$({\bf x}_{\rm obs},z_{\rm obs},\nu_{\rm obs})$ from a source of 
differential luminosity $L_\nu$ emitted at 
$({\bf x}_{\rm s},z_{\rm s},\nu_{\rm s})$ is given by
\begin{equation}
F_{\nu,s}({\bf x}_{\rm obs},z_{\rm obs},\nu_{\rm obs}) = 
\frac{L_{\nu}\left(\nu=\nu_{\rm s}\right)}{4\pi D_L^2(z_{\rm obs},z_{\rm s})}
\cdot\left(\frac{1+z_{{\rm s}}}{1+z_{{\rm obs}}}\right) 
\cdot \exp\left[-\tau_{\nu_{\rm obs}} \right],
\label{eq:flux2}
\end{equation}
where the factor $(1+z_{\rm s})/(1+z_{\rm obs})$ reflects the fact that 
the observer sees the differential frequency interval reduced by redshift
relative to the emitted interval.
Here $D_L(z_{\rm obs},z_{\rm s})$ is the luminosity distance given by
\begin{equation}
D_L(z_{\rm obs},z_{\rm s})\equiv \left(\frac{r_{\rm os}}{1+z_{\rm obs}}\right)
\left(\frac{1+z_{\rm s}}{1+z_{\rm obs}}\right),
\end{equation}
where the factor $(1+z_{\rm s})/(1+z_{\rm obs})$ takes account of the 
redshifting of photon energies and arrival rates, which reduce the 
observed flux by $[(1+z_{\rm obs})/(1+z_{\rm s})]^2$.
The optical depth $\tau_{\nu_{\rm obs}}$ depends on the observed
frequency $\nu_{\rm obs}$ according to 
\begin{equation}
\tau_{\nu_{\rm obs}}
=\int_{t(z_{\rm s})}^{t(z_{\rm obs})}\rho_b({\bf x},z)\kappa_{\nu'}({\bf x},z)cdt,
\label{equ:tau-general}
\end{equation}
where $\nu'=\left(\frac{1+z}{1+z_{\rm obs}}\right)\nu_{\rm obs}$, 
$\rho_b({\bf x},z)$ is the baryon density at $({\bf x},z)$, 
$\kappa_{\nu'}({\bf x},z)$ is the opacity at $({\bf x},z)$ to photons 
of frequency $\nu'$, and where ${\bf x}$ and z are the position and redshift
of photons travelling along the line of sight which were emitted at 
$({\bf x}_{\rm s},z_{\rm s})$ and will be received at 
$({\bf x}_{\rm obs},z_{\rm obs})$.

The optical depth of the IGM to continuum UV photons in the LW range
between 11.2 eV and 13.6 eV is predominantly due to resonant absorption 
by neutral H atoms in the Lyman series lines with upper states $n=i$, 
$i\geq3$ (HAR). The mean optical depths in these lines can be written as 
follows:
\begin{equation}
\tau_i=\frac{f_{\rm osc,i}}{f_{\rm osc,\alpha}}\frac{\nu_\alpha}{\nu_i}
\tau_{\rm GP,\alpha}
=\left(\frac{f_{\rm osc,i}}{0.416}\right)\left(\frac{0.75}{1-1/i^2}\right)
\tau_{\rm GP,\alpha}
\rightarrow1.8f_{\rm osc,i}\tau_{\rm GP,\alpha}\,\,\,\,\,{\rm for\,\,
  i\gg1},
\label{eq:taui}
\end{equation}
where $f_{\rm osc,i}$ and $f_{\rm osc,\alpha}$ are the oscillator strengths
and $\nu_i$ and $\nu_\alpha$ are the frequencies, for lines of upper states 
$n=i$ and 2, respectively, and where $\tau_{\rm GP,\alpha}$ is the familiar 
Gunn-Peterson optical depth in the Lyman-$\alpha$ transition:
\begin{eqnarray}
\tau_{\rm GP,\alpha}&=&\left(\frac{\pi e^2}{m_ec}\right)
      \left(\frac{f_{\rm osc,\alpha}}{\nu_\alpha}\right)
      \left(\frac{c}{H(z)}\right)n_{\rm H\,I}\nonumber\\
&=&2.2\times10^6\left(\frac{\Omega_b}{0.044}\right)\left(\frac{h}{0.7}\right)^{-1} 
\left(\frac{\Omega_m}{0.27}\right)^{-1/2}
x_{\rm H\,I}\left(\frac{1+z}{21}\right)^{3/2},
\label{eq:tauGP}
\end{eqnarray}      
where $n_{\rm H\,I}$ is the mean number density of neutral hydrogen
and $x_{\rm H\,I}$ is the mean neutral fraction of the IGM. Since we are  
most interested here in the early phases of the EOR, 
$x_{\rm H\,I}=1-x_{\rm H\,II}\approx1$. As such, 
$\tau_{\rm GP,\alpha}\approx10^6\gg1$. Even {\it inside} H~II regions, 
however, $x_{\rm H\,I}\gtrsim10^{-4}$ \citep{2007arXiv0711.2944I}, so  
$\tau_{\rm GP,\alpha}\gg1$ in general.  As $i$ increases, $f_{\rm osc,i}$ 
decreases (e.g. $f_{\rm osc,i}=0.079,$ 0.029, and 0.014 for $i=3,$ 4, and 5,
respectively \citep[e.g.][]{fosc_citation}, so there is some $i_{\rm max}$ 
such that $\tau_i<1$ for $i>i_{\rm max}$. 
For example, if {\em mean} IGM density is assumed in the ionized
region with $x_{\rm HI}\simeq 10^{-4}$, 
$i_{\rm max}=8$, 7, and 6 
with $f_{\rm osc}=0.0032$, $0.0048$,
and $0.0078$ at $z=20$, 15,
and 10, respectively. Note, however, that cosmic reionization occurs
in an inside-out fashion, such that overdense regions are 
ionized earlier than the mean or underdense regions
(e.g. \citealt{2007MNRAS.376..534I}). The effective $i_{\rm max}$ in
ionized regions, therefore, can be much larger than the estimates above.
According to HAR, $i_{\rm max}\sim150$
in the neutral IGM. It is a good aproximation, therefore, to assume that the
Lyman lines are optically thick for all $i$, since the frequency range 
over which the lines are not optically thick is a negligible fraction of the
LW range below 13.6 eV.

The opacity of the IGM due to LW band absorption by ${\rm H_2}$ is relatively 
unimportant by comparison, since the ${\rm H_2}$ concentration in the IGM is
small even before the LW background rises to suppress it. The pre-reionization 
${\rm H_2}$ concentration is $\sim10^{-6}$ \citep{1994ApJ...427...25S}, and 
so $\tau_{\rm LW}<1$. We shall, henceforth, neglect this source of
opacity. We refer the reader to \S \ref{sec:conclusions} for more
rigorous justification. 

The expected number of computational operations required to evaluate
equations~(\ref{eq:bruteforce}) and (\ref{eq:flux2}) is
$N_{\rm s}\cdot N_{\rm g}\cdot N_{\rm f}$, where $N_{\rm s}$ is
the number of sources, $N_{\rm g}$ is the number of grid cells on
which $J_\nu$ is calculated, and $N_{\rm f}$ is the number of
frequency-bins which are required to resolve the frequency dependence 
of $\tau_{\nu_{\rm obs}}$ adequately. While these equations are 
straightforward, evaluating them numerically in a brute-force way 
can be prohibitively expensive in computational resources. For example, 
the H Lyman series resonance lines are optically thick to photons in
the LW bands frequency range, 
and one should consider $N_{\rm f}\gg 1$ to account for this effect 
properly. Currently, a full 3D, multi-frequency radiative transfer 
calculation is not feasible for the problem of interest. The effective 
number of sources $N_{\rm s}$ in our problem can be as large as $10^7$ 
due to the large size of the LW horizon, and we need about 
$N_{\rm g}\gtrsim 10^6$ grid-points to produce a statistically 
significant result. Cosmic reionization simulations, by contrast, 
do not require a multi-frequency operation, and yet, these simulations 
have only just become feasible recently with the help of 
massively-parallel computers.

The following sections describe how we overcome this technical difficulty
in calculating the LW background by reducing $N_{\rm f}$ to 1, even
though the net result becomes equivalent to a full multi-frequency
radiative transfer calculation. We further describe in detail how we
sum individual $F_{\nu,\,{\rm s}}$'s, taking full account of the effect of
redshifting and of the finite light-crossing time between sources and 
observers.

\subsection{\label{sub:picket-fence}Attenuation of ${\rm H}_{2}$ Dissociating
Photons from a Single Source: the ``Picket-Fence'' Modulation Factor}

\begin{figure}[ht]
\includegraphics[width=0.5\textwidth]{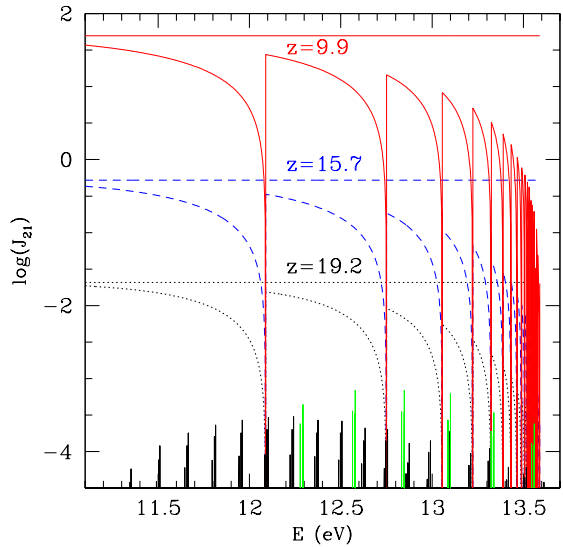}
\caption{The observed ``sawtooth'' modulation of the uniform, isotropic
radiation background observed in the UV range of the LW band range for 
a homogeneous $\Lambda$CDM universe with flat-spectrum sources, caused 
by H Lyman line opacity of the IGM (see text), for redshifts $z=19.2$ 
(bottom lines, black), $z=15.7$ (middle lines, blue) and $z=9.9$ (top 
lines, red). The horizontal (dashed, corresponding colors) lines show 
the unattenuated mean
intensity levels at these redshifts. A spatially uniform emissivity is
assumed which evolves in time in proportion to the collapsed fraction
of the matter density in halos massive enough to be sources of 
reionization, as described in \S~\ref{sec:applic}. 
Also plotted in vertical 
lines are the locations of relevant LW bands, when ${\rm H}_2$ are 
assumed to be in the ground electronic state $X{}^{1}\Sigma_{g}^{+}$
with $v''=0$ and $J''=0,\,1$. The height of these lines
corresponds to $\log(0.01\times f_{\rm osc})$, where $f_{\rm osc}$ is
the oscillator strength of Lyman (black) and Werner (green) bands
compiled by \citet{1989A&AS...79..313A}.
\label{fig:sawtooth}}
\end{figure}

For an inhomogeneous distribution of sources, we need to calculate the 
attenuation of continuum photons emitted in the LW energy range 
11.2 - 13.6 eV separately for each individual source, by hydrogen Lyman 
line resonance absorption and subsequent cascades along the light cones 
from the source to every observer. Consider a source emitting
continuum radiation at frequency $\nu_s$ at redshift $z=z_{{\rm s}}$.
As the photon travels toward the observer, it is absorbed when its 
frequency redshifts into an H Lyman series resonance line and, some of the 
time, the decay of the excited state replaces the original photon with
photons at frequencies below the range of the LW bands. If the original
photon is resonantly scattered, it is quickly reabsorbed, until all 
resonant photons eventually turn into low-frequency photons below the LW 
bands. For this reason, \citet[HRL, hereafter]{1997ApJ...484..985H} and 
HAR assumed that, whenever the photons emitted in this range redshifted 
into one of the H Lyman resonance lines, they were completely attenuated 
and turned into low-frequency photons --- mostly Ly$\alpha$ photons --- 
out of the LW range. From the observer's viewpoint, this leads to a 
series of {}``dark screens'', defined as sharp boundaries beyond which 
the observer cannot see any sources contributing LW intensity.
These are marked by the maximum redshifts $z_{{\rm max}}$ defined by 
\begin{equation}
\frac{1+z_{{\rm max}}}{1+z_{{\rm obs}}}=\frac{\nu_{i}}{\nu_{{\rm obs}}},
\label{eq:rshift_Haiman}
\end{equation}
 where $\nu_{i}$ is the frequency of the Lyman line closest to the 
observed frequency $\nu_{{\rm obs}}$ from above. For a {\it homogeneous} 
universe with spatially-uniform emissivity, this results in the well-known 
{}``sawtooth'' modulation of a uniform, isotropic LW background spectrum 
(HRL; HAR), an example of which we have plotted in Figure~\ref{fig:sawtooth}
for a homogeneous $\Lambda$CDM universe with flat-spectrum sources.

We shall make the same assumption here, that all LW photons are completely 
attenuated once they redshift into an H Lyman resonance line with upper 
state $n\geq3$. However, we cannot limit ourselves to the homogeneous 
universe approximation. Since our objective is to consider contributions 
from individual sources that are distributed inhomogeneously, we must, 
instead, calculate how continuum photons emitted by {\it each} source 
will be attenuated by hydrogen atoms which they encounter along the 
particular line of sight that connects them with a given point of 
observation. We shall describe the attenuation of an individual source
here in what follows. We shall then describe in \S~\ref{sub:summing} 
how we sum over these individual source contributions to obtain the 
spatially-varying LW background intensity. 

For the homogeneous universe in which the observed spectrum is 
transformed by the sawtooth modulation shown in Figure~\ref{fig:sawtooth},
this spectrum is the result of superposing the spectra of sources 
distributed continuously in lookback time along the line of sight.
In that case, a given observed frequency combines the effect of photons 
emitted at different lookback times which experience different amounts
of redshifting before reaching the observer. It is natural, then, to 
describe the modulation from the viewpoint of the observer, with the 
edge of each saw-tooth corresponding to the ``dark screens'' at the 
frequencies of the H Lyman lines, and the spectrum to the red side of 
a given line originating in the past from sources nearer than the 
distance to the corresponding screen. When we consider, instead, the 
spectrum of a single source in an inhomogeneous universe, there are
also ``dark screens'' beyond which LW photons cannot pass, but these 
screens are best described from the point of view of the source. In 
that case, as a photon travels from the source, it will survive only 
until it encounters a ``dark screen'' in its future, as it redshifts 
into the nearest Lyman line. From the viewpoint of a source, the dark 
screens are located at the maximum radii that photons emitted at 
different frequencies can travel from a source. Instead of 
equation~(\ref{eq:rshift_Haiman}), these radii are marked
by minimum redshifts $z_{{\rm min}}$ defined by 
\begin{equation}
\frac{1+z_{{\rm min}}}{1+z_{{\rm s}}}=\frac{\nu_{i}}{\nu_{{\rm s}}},
\label{eq:rshift-jin}
\end{equation}
 where $\nu_{i}$ is the (observed) frequency of the Lyman line closest
to the emitted frequency $\nu_{{\rm s}}$ from below. A screen corresponding 
to $\nu_{i}$ has the radius $r_{i}(\nu_{{\rm s}})$, where 
$r_{i}(\nu_{{\rm s}})\equiv r_{{\rm os}}(\nu_{i};\,\nu_{{\rm s}},\, z_{{\rm s}})$.
This introduces finite frequency gaps in
the transmitted spectrum, which as a consequence resembles a ``picket 
fence'', with the intensity unattenuated between the dark gaps, as 
follows.

\begin{figure}[ht]
\includegraphics[width=0.5\textwidth]{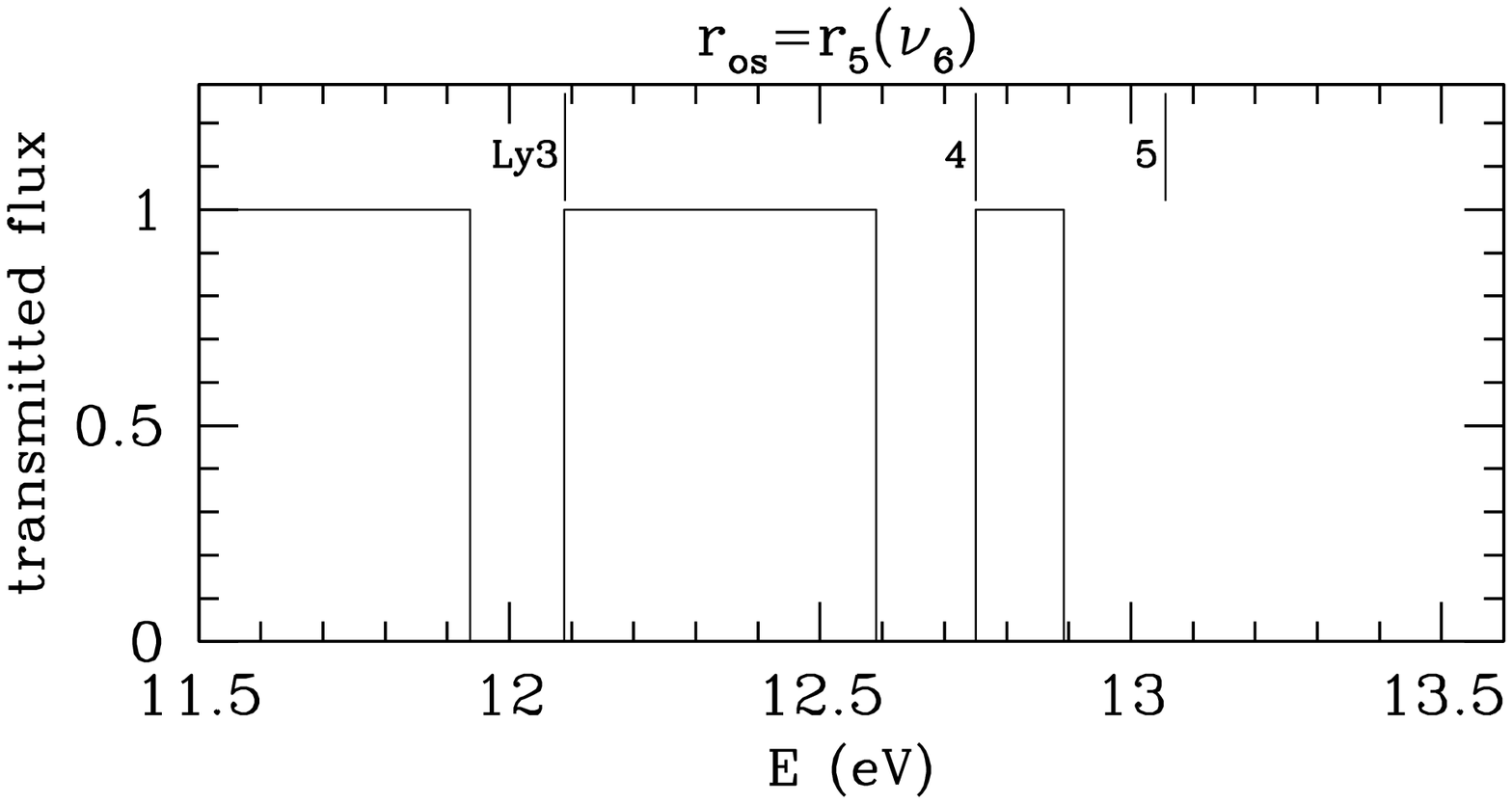}
\includegraphics[width=0.5\textwidth]{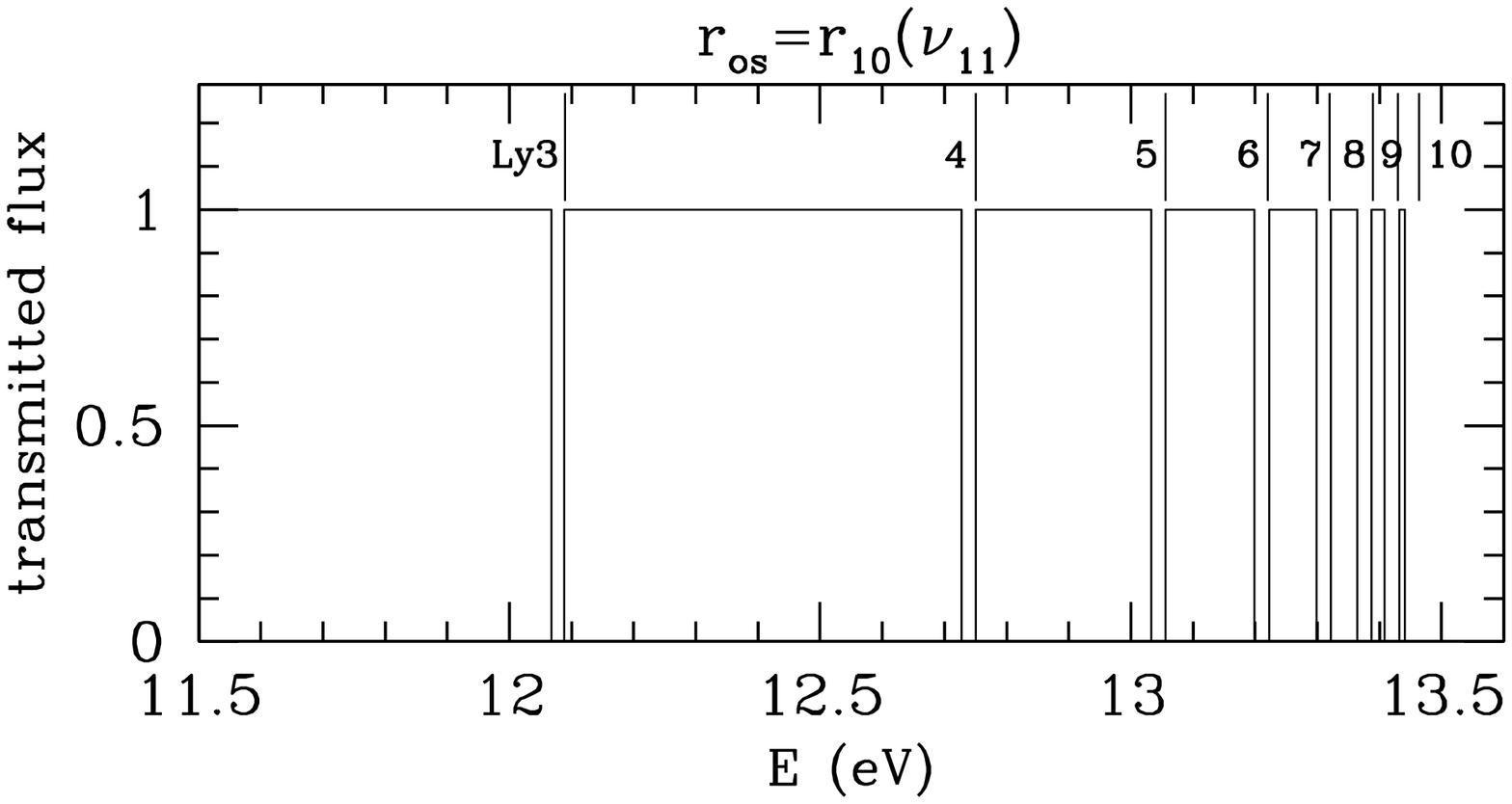}
\caption{Transmitted spectrum from a single source: picket-fence modulation.
({\em left}) Relative flux observed at $r_{5}(\nu_{6})$, where a
photon emitted at Ly5 (Ly$\epsilon$) frequency is redshifted into Ly4 
(Ly$\delta$)
frequency. ({\em right}) Relative flux observed at $r_{10}(\nu_{11})$,
where a photon emitted at Ly$11$ frequency is redshifted into Ly$10$
frequency. We use a convention that the energy of Ly$i$ photon is
$13.6\,{\rm eV}\,\left(1-1/i^{2}\right)$. The loocation of several
Lyman resonance lines Ly$i$ is shown in vertical lines and denoted by
$i$ in both panels. 
\label{fig:picketfence}}
\end{figure}

Consider a dark screen located at comoving distance $r_{i}(\nu_{i+1})$ 
from the source defined by 
\begin{equation}
r_{i}(\nu_{i+1})\equiv r_{{\rm os}}(\nu_{i};\,\nu_{i+1},\, z_{{\rm s}}),
\label{eq:rii}
\end{equation}
where a photon emitted at $\nu_{i+1}$ from a source is redshifted into 
$\nu_{i}$. At this location, all the photons with $\nu_{{\rm s}}\ge\nu_{i}$
are completely attenuated in the following way. First, a photon with
$\nu_{{\rm s}}=\nu_{i+1}$ will be redshifted into $\nu_{i}$ and
attenuated at $r_{i}(\nu_{i+1})$%
\footnote{A more accurate description is that a photon with frequency slightly
smaller than $\nu_{i+1}$ is redshifted into $\nu_{i}$ and attenuated
at distance slightly shorter than $r_{i}(\nu_{i+1})$, because a photon
with $\nu_{s}=\nu_{i+1}$ is attenuated on-site right after being
emitted. However, we give the description in the text for simplicity.%
}. Any photons with $\nu_{{\rm s}}$ in between $\nu_{i}$ and $\nu_{i+1}$
will then be redshifted into $\nu_{i}$ and attenuated at some distance
shorter than $r_{i}(\nu_{i+1})$. Since no photon can cross a Lyman line 
frequency as it is redshifted, the observed spectrum will be completely 
black inside a trough between $\nu_{i}$ and $\nu_{i+1}$. For 
$\nu_{i+1}\le\nu_{{\rm obs}}<\nu_{i+2}$,
this will happen at $r_{i+1}(\nu_{i+2})$. Because 
\begin{equation}
r_{i+1}(\nu_{i+2})<r_{i}(\nu_{i+1}),
\label{eq:riiri}
\end{equation}
 the observed spectrum at $r_{i}(\nu_{i+1})$ will also have a trough
from $\nu_{i+1}$ to $\nu_{i+2}$. This way, all the photons with
$\nu_{{\rm obs}}\ge\nu_{i}$ are completely attenuated at 
$r_{{\rm os}}=r_{i}(\nu_{i+1})$.
On the other hand, because $r_{i-1}(\nu_{i})>r_{i}(\nu_{i+1})$, a
photon with $\nu_{{\rm s}}=\nu_{i}$ has not fully redshifted into
$\nu_{i-1}$ but only into $\nu_{{\rm obs}}(r_{i}(\nu_{i+1});\,\nu_{i},\, z_{{\rm s}})$
at $r_{i}(\nu_{i+1})$, where 
\begin{equation}
\nu_{{\rm obs}}(r_{{\rm os}};\,\nu_{{\rm s}},\, z_{{\rm s}})
\equiv\nu_{{\rm s}}\left[1+\frac{H_{0}\Omega_{m}^{1/2}}{2c}(1+z_{{\rm s}})^{1/2}r_{{\rm os}}\right]^{-2},
\label{eq:nuzr}
\end{equation}
obtained from equation (\ref{eq:r_os2}). Photons with the observed
frequency ranging from $\nu_{i-1}$ to 
$\nu_{{\rm obs}}(r_{i}(\nu_{i+1});\,\nu_{i},\, z_{{\rm s}})$ will have 
reached $r_{i}(\nu_{i+1})$ without attenuation. Therefore, the spectrum 
will have full transmission for 
$\nu_{i-1}<\nu_{{\rm obs}}<\nu_{{\rm obs}}(r_{i}(\nu_{i+1});\,\nu_{i},\, z_{{\rm s}})$,
while there is a completely black trough for 
$\nu_{{\rm obs}}(r_{i}(\nu_{i+1});\,\nu_{i},\, z_{{\rm s}})\le\nu_{{\rm obs}}<\nu_{i}$.
Similarly, the next lower-energy interval, defined by 
$\nu_{i-2}<\nu_{{\rm obs}}<\nu_{i-1}$, will have full transmission for 
$\nu_{i-2}<\nu_{{\rm obs}}<\nu_{{\rm obs}}(r_{i}(\nu_{i+1});\,\nu_{i-1},\, z_{{\rm s}})$
and a trough for $\nu_{{\rm obs}}(r_{i}(\nu_{i+1});\,\nu_{i-1},\, z_{{\rm s}})
\le\nu_{{\rm obs}}<\nu_{i-1}$. As a result, the observed relative flux is 
affected by what we call the {}``picket-fence'' modulation, as depicted in 
Figure~\ref{fig:picketfence}.

Note that, at a fixed distance, $\nu_{\rm obs}\propto\nu_{s}$ from equation
(\ref{eq:nuzr}). A gap appearing between observed frequencies $\nu_{j}$
and $\nu_{j+1}$ at some $r_{{\rm os}}$ is also proportional to $\nu_{j+1}$.
If we call the size of this gap $\Delta\nu_{{\rm gap},\, j+1}$, we
then have 
\begin{equation}
\frac{\Delta\nu_{{\rm gap},\, j+1}}{\Delta\nu_{{\rm gap},\, i+1}}
=\frac{\nu_{j+1}}{\nu_{i+1}},
\label{eq:delnu}
\end{equation}
 as long as the range of frequency, $[\nu_{j},\,\nu_{j+1}]$, is not
fully covered by $\Delta\nu_{{\rm gap},\, j+1}$. 

We define the picket-fence modulation factor $f_{{\rm mod}}$ by
\begin{equation}
f_{{\rm mod}}(r_{{\rm os}};\, z_{{\rm s}})
      \equiv\langle\exp(-\tau_{\nu_{\rm obs}})\rangle
      \equiv \frac{\int_{11.5\,{\rm eV}}^{13.6\,{\rm eV}}d(h\nu_{\rm obs})\,
  \exp(-\tau_{\nu_{\rm obs}})}{\int_{11.5\,{\rm eV}}^{13.6\,{\rm eV}}d(h\nu_{\rm obs})},
\end{equation}
 which is a function only of the comoving distance $r_{{\rm os}}$
and the source redshift $z_{{\rm s}}$. We choose 11.5 eV as the minimum
energy of interest, because the dissociation rate for the LW bands at 
$h\nu<11.5\,{\rm eV}$ is negligible compared to that for those at 
$h\nu>11.5\,{\rm eV}$ (see, e.g., Figure 1 of HAR). When ${\rm H_2}$ molecules 
absorb photons in the LW bands, in general only about 
15\% of these excitations lead to dissociation of ${\rm H}_{2}$. When a 
single source is observed at some comoving distance $r_{{\rm os}}$, some 
LW bands will be excited by fully transmitted photons, which results in 
dissociation about 15\% of the time, therefore, while other bands will 
not, because they reside in a trough. For sources of interest here, we 
can approximate $L_{\nu}$ by a flat spectrum whose amplitude is given by 
the frequency-averaged luminosity $\left\langle L_{\nu}\right\rangle 
\equiv\int_{11.5\,{\rm eV}}^{13.6\,{\rm eV}}d(h\nu_{{\rm s}})\, 
          L_{\nu}(\nu_{{\rm s}})/(2.1\,{\rm eV})$.
This is a fairly good approximation, unless the spectrum is unusually 
steep in the narrow energy range of interest, 
$[11.5\,-\,13.6]\,{\rm eV}$. For example, both black-body spectra with 
$T=[5-10]\times 10^4 \,{\rm K}$ and a power-law spectrum with 
$L_{\nu}\propto \nu^{-1}$ have a maximum deviation from 
$\left\langle L_{\nu}\right\rangle$ which is smaller than 10\% in this 
energy range. In that case, because the LW bands are almost uniformly 
distributed in frequency, the true dissociation
rate will be almost identical to that obtained by assuming that all
the ${\rm H}_{2}$ LW lines experience the frequency-averaged
value of the LW intensity after the picket-fence modulation. Therefore,
in addition to the geometrical dilution of the incident flux, the 
${\rm H}_{2}$ dissociation rate will be suppressed in proportion to 
$f_{\rm mod}$. This $f_{\rm mod}$ is just the fraction of the total 
frequency interval from 11.5 to 13.6 eV observed at $z_{\rm obs}$ from 
a source at $z_{\rm s}$ at comoving distance $r_{\rm os}$ occupied by 
the full transmission windows in between the dark troughs, as described
above. Hence, 
\begin{equation}
f_{\rm mod}=1-\sum_j\left(\frac{h\Delta\nu_{\rm gap,j}}{2.1 \,\rm eV}\right).
\end{equation}

\begin{figure}[ht]
\includegraphics[width=0.5\textwidth]{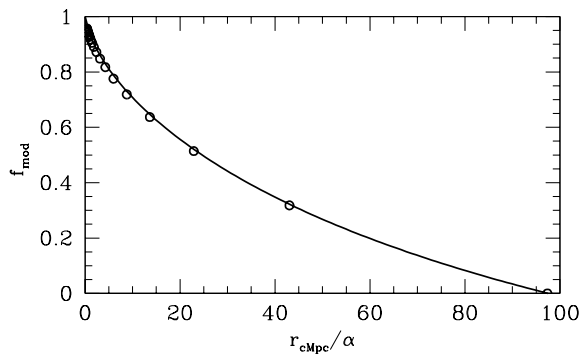}
\caption{Picket-fence modulation factor $f_{{\rm mod}}$ as a function
  of comoving distance
$r_{{\rm cMpc}}$ in units of Mpc. True values at selected radii $r_i
  (\nu_{i+1})$ 
  (open circle) and a fit (solid curve) are plotted. $\alpha$, given
  by equation 
  (\ref{eq:scaling}), 
is a distance scaling factor which depends on redshift. When $r_{{\rm
  cMpc}}=97.39\alpha$, 
$f_{{\rm mod}}=0$, which sets the {}``LW horizon'' for ${\rm H}_{2}$
dissociating radiation from a source.\label{fig:fmod}}
\end{figure}

The picket-fence modulation factor is a key ingredient in alleviating
computational difficulties which would have arisen due to a 
multi-frequency calculation. We have calculated $f_{{\rm mod}}$ numerically
and found a simple fitting formula which fits the true values within a 2\% 
error (see Figure \ref{fig:fmod}): 
\begin{equation}
f_{{\rm mod}}=\cases{
1.7\,\exp\left[-(r_{{\rm cMpc}}/116.29\alpha\right)^{0.68}]-0.7 & 
if $r_{\rm cMpc}/\alpha \le 97.39$ \\
0 & if $r_{{\rm cMpc}}/\alpha>97.39$}
\label{eq:fmod-fitting}
\end{equation}
where $r_{{\rm cMpc}}$ is $r_{{\rm os}}$ in units of comoving Mpc,
and $\alpha$ is a scaling factor given by 
\begin{equation}
\alpha=\left(\frac{h}{0.7}\right)^{-1}
\left(\frac{\Omega_{m}}{0.27}\right)^{-1/2}
\left(\frac{1+z_{{\rm s}}}{21}\right)^{-1/2}.
\label{eq:scaling}
\end{equation}
We call $r_{3}(\nu_{4})$, which is equal to $97.39\,\alpha\,{\rm cMpc}$,
the {}``LW horizon'' $r_{\rm LW}$. This is the maximum comoving distance 
from a source that an ${\rm H}_{2}$ dissociating photon can reach, 
corresponding to the distance from which the redshift produces the 
maximum frequency difference possible between two adjacent lines in the 
Lyman series (as long as we restrict the observed energy range to 
[11.5 - 13.6] eV). Note that all the dark screen distances are scaled 
by $\alpha$, which increases as $z_{{\rm s}}$ decreases.

\subsection{\label{sub:summing}Intensity of ${\rm H}_{2}$ dissociating 
photons from multiple sources}

\begin{figure}[t]
\includegraphics[width=0.5\textwidth]{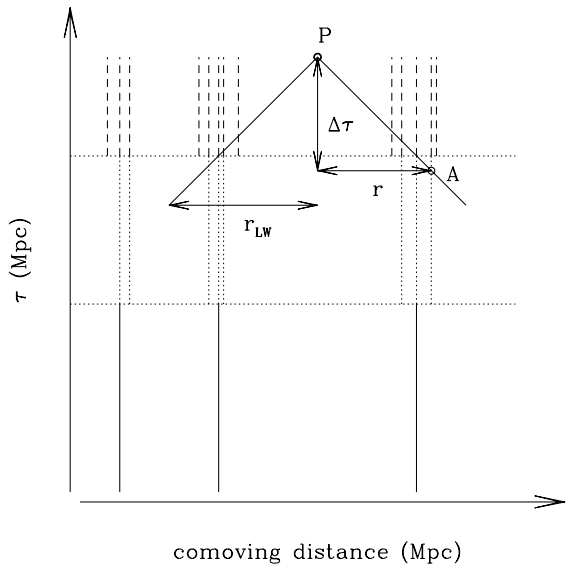}

\caption{Conformal space-time diagram of radiation sources and the 
past light cone of an observer, used to identify which of the sources
in the N-body simulation volume in the past emitted light just now 
reaching the observer at a given time. Radiation sources are created 
discretely in time in the N-body simulation results --- i.e. source 
catalogue is constructed at each output time (dotted horizontal lines). 
The location of each source is assumed to stay constant during each time 
step (shown as solid, dotted, and short-dashed vertical lines) as are
the source luminosities. Photons follow null geodesics 
truncated at the LW horizon $r_{\rm LW}$, and an observer at point P 
will see sources whose world lines intersect with the past light-cone. 
The flux contributed by these sources are determined by where these 
intersecting points lie along the time axis. We use a coordinate system 
composed of comoving distance and conformal time, for computational ease. 
For example, the conformal lookback time $\Delta\tau$ (Mpc) from point P 
to a source at point A, which determines the emitting redshift and the 
source flux, is easily obtained once the comoving distance $r$ (Mpc) to 
the source is known, because $\Delta\tau=r$.\label{fig:world}} 
\end{figure}

The path of a photon in the expanding universe follows a null geodesic.
The Friedmann-Robertson-Walker metric for a homogeneous, isotropic 
universe is given by
\begin{equation}
ds^{2}=dt^{2}-a^{2}(t)\left[dr^{2}+r^{2}d\Omega^{2}\right]
=a^{2}(t)\left[d\tau^{2}-\left(dr^{2}+r^{2}d\Omega^{2}\right)\right],
\label{eq:frw}
\end{equation}
where we adopt natural units with $c=1$. In that case, the comoving 
distance travelled by a photon since its emission is given by setting 
$ds^2=0$ in equation~(\ref{eq:frw}), solving for $dr$ ($d\Omega=0$), 
and integrating over time to yield the conformal time
$\tau$, defined by 
\begin{equation}
\tau=\int d\tau=\int dt/a(t).
\label{eq:conformal}
\end{equation}
We use this fact to construct the world lines of sources and their 
radiation, showin in Figure~\ref{fig:world}. If our choice of 
space-time coordinates is the comoving 
distance and the conformal time, then null geodesics make straight 
lines at a $45^{\circ}$ angle. World lines of the sources, on the 
other hand, will be close to straight lines, parallel to the 
conformal time axis. For simplicity, we will neglect the small 
peculiar motion of halos.

We must account for the finite light-crossing time for light from
sources to reach an observer, because these are distributed over a
truly cosmological volume and the population of sources can vary 
significantly over the lookback time
corresponding to $r_{\rm LW}\sim 100 {\rm cMpc}$, due to the rapid
evolution of cosmological structure. The conformal space-time diagram
of sources mentioned above becomes a useful tool for this task.  
At a given redshift we draw past light cones from an observing point, 
which have a maximum length equal to the LW horizon length, $r_{\rm LW}$. 
When the world line of a source intersects one of these past light cones, 
we add its flux contribution to the mean intensity at the corresponding
observing point (see Figure~\ref{fig:world}). The fact that 
$\Delta \tau=r_{\rm os}$, where $\Delta \tau$ is the conformal lookback
time to a source at comoving distance $r_{\rm os}$ from the observing
point, makes it easy to find these intersecting points as well as
the source redshift. The conformal time interval, $\tau_{\rm LW}$, which 
corresponds to the LW horizon, $r_{\rm LW}$, determines how far back in 
look-back time a given observer cell at a given epoch $z_{\rm obs}$ must
extend its past light-cone to look for contributing sources. Accordingly,
this operation requires that the past light-cones extend back through a
number, $n_{\rm steps}$, of time steps, $\Delta t$, equal to 
$\tau_{\rm LW}/\Delta t$. 

After a contributing source is found, its frequency-averaged LW flux 
observed at the given ${\bf x}_{\rm obs}$ and $z_{\rm obs}$ is evaluated 
using equation~(\ref{eq:flux2}), replacing $L_\nu$ by its average over
the LW band frequency and replacing $\exp(-\tau_{\rm obs})$ by $f_{\rm mod}$
using equations~(\ref{eq:fmod-fitting}) and (\ref{eq:scaling}).
We sum fluxes from all the sources (denoted by the subscript s)
 observed within the LW horizon.

\section{The Inhomogeneous LW Background from a Simulation of Cosmic 
Reionization
\label{sec:applic}}

\subsection{\label{sub:reion_sim}Illustrative Case: Self-Regulated 
Reionization}

As an example, we apply the methodology for calculating the fluctuating 
LW background described in the previous sections to one of our large-scale 
N-body and radiative transfer simulations of cosmic reionization presented
in \citet[henceforth ``IMSP'']{2007MNRAS.376..534I}. The cosmological structure 
formation and evolution is followed with a particle-mesh N-body code called
PMFAST (\citealt{2005NewA...10..393M}). These N-body results then provide 
the evolving density field  of the IGM (coarsened to a lower resolution
than the original particle-mesh grid in order to make the radiative
transfer feasible) and the location and mass of all the
halo sources, as input to a separate radiative transfer simulation of
inhomogeneous reionization. The latter simulation is performed by our
${\rm C}^2-$Ray ({\bf C}onservative, {\bf C}ausal {\bf Ray}-Tracing)
code, a grid-based, ray-tracing, radiative transfer and nonequilibrium
chemistry code, described in \citet{methodpaper}. The ionizing
radiation is ray-traced from every source to every grid cell at a given
timestep using a method of short characteristics. The code is explicitly
photon-conserving in both space and time, which ensures an accurate
tracking of ionization fronts, independent of the spatial and time
resolution, even for grid cells which are optically thick to ionizing
photons and time steps long compared to the ionization time of the
atoms, with correspondingly great gains in efficiency. The code has been
tested against analytical solutions (\citealt{methodpaper}) and, in
direct comparison with other radiative transfer methods, on a
standardized set of benchmark problems
(\citealt{2006MNRAS.371.1057I}).

We simulated the $\Lambda$CDM universe with $1624^3$ dark matter
particles of mass $10^6 \,{\rm M}_{\odot}$, in a comoving simulation
volume of $(35\, h^{-1}\,{\rm Mpc})^3$. This allowed us to resolve
(with 100 particles or more per halo) all halos with mass of 
$10^8 \,{\rm M}_{\odot}$ or above. This is roughly the minimum mass of 
halos which can radiatively cool by hydrogen atomic-line excitation and
efficiently form stars. The radiative transfer grid has $203^3$ cells.

\begin{figure*}[ht]
\includegraphics[width=0.5\textwidth]{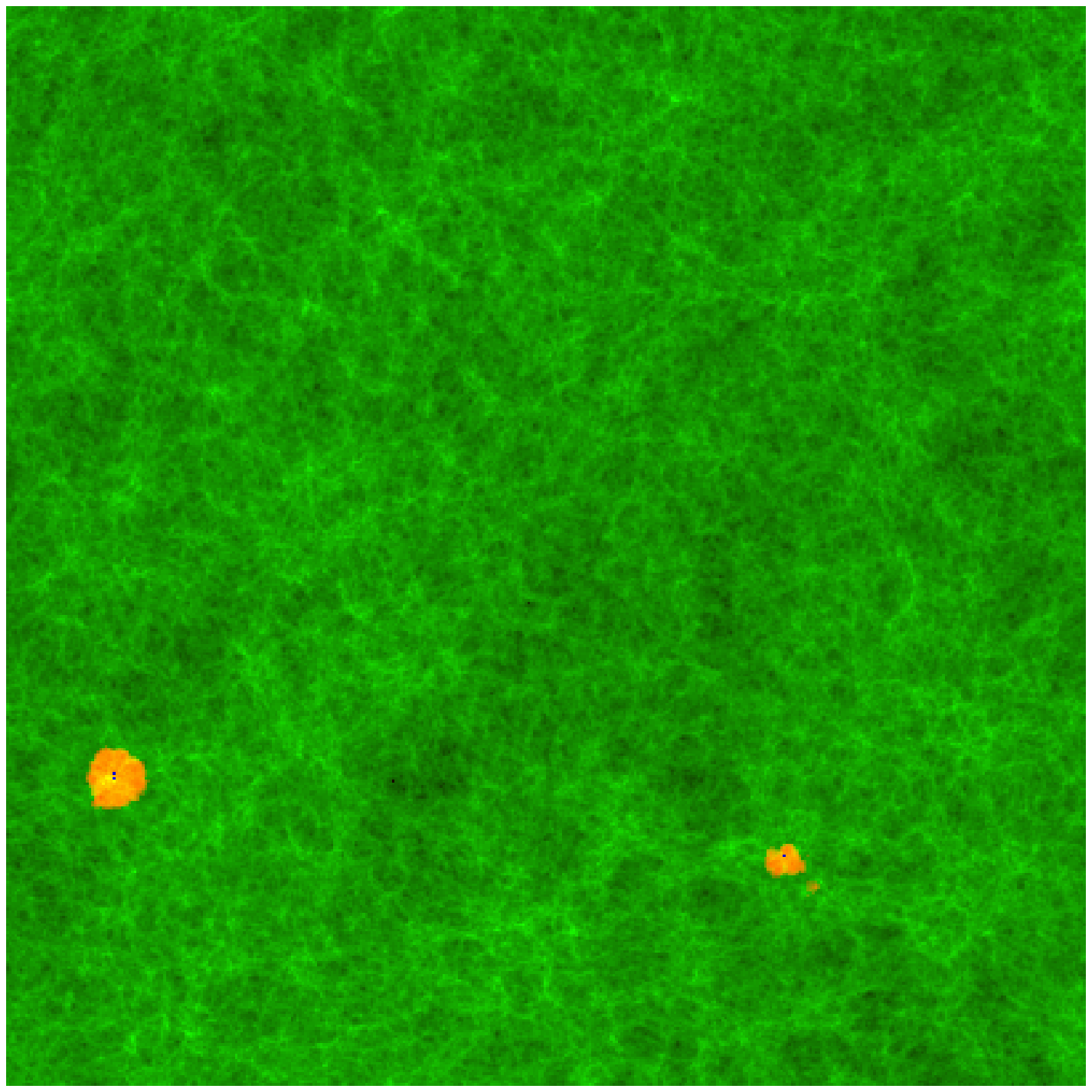}
\includegraphics[width=0.5\textwidth]{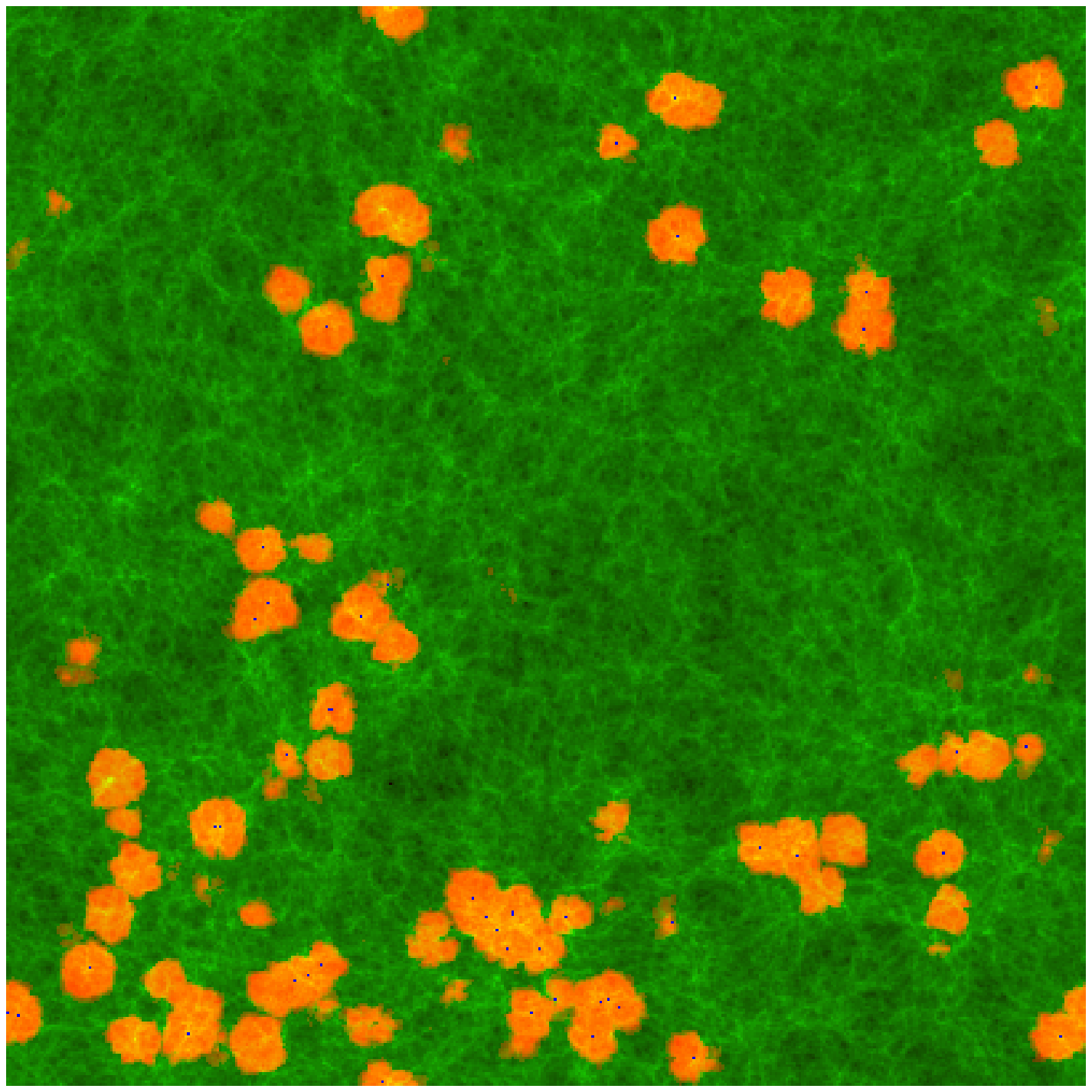}

\includegraphics[width=0.5\textwidth]{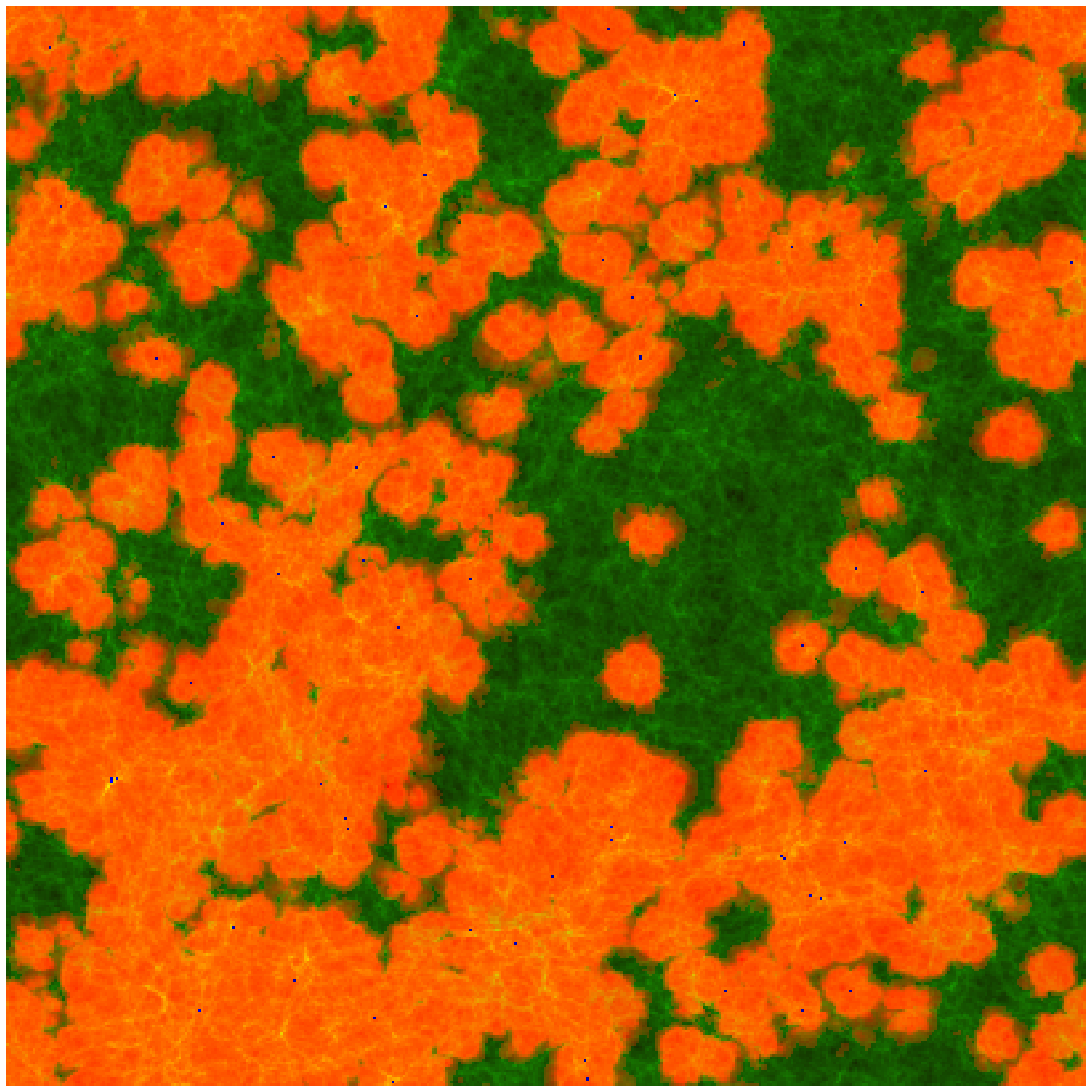}
\includegraphics[width=0.5\textwidth]{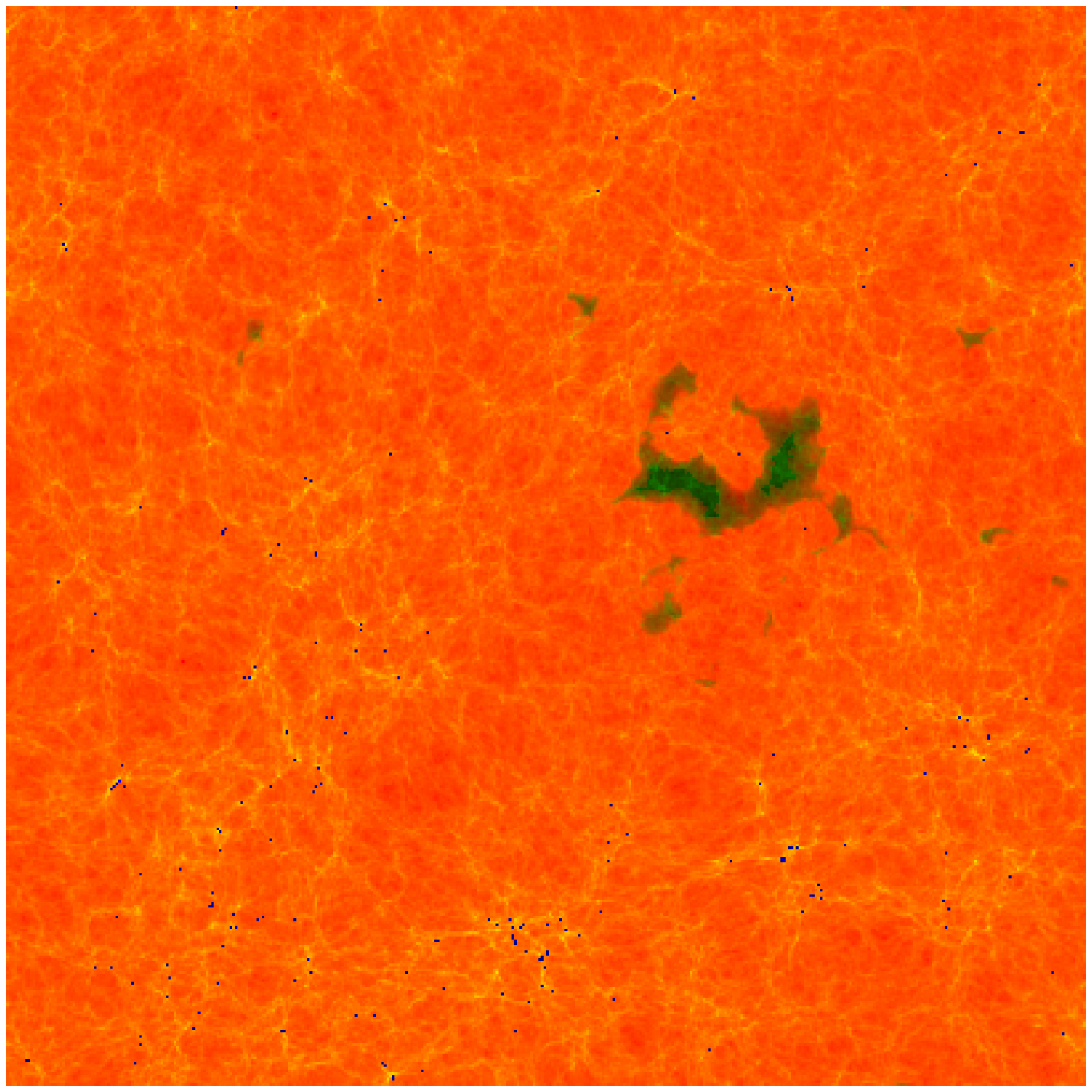}
\caption{Spatial slices of the ionized and neutral gas density from
 self-regulated simulation (IMSP; f2000\_250S case), at redshifts
 $z=15.7$ (top-left), 12.9 (top-right), 9.9 (bottom-left), and 7.9
 (bottom-right). The volume-weighted (mass-weighted) global ionized
 fractions at these redshifts are $6.4\times 10^{-3}$ ($9.7\times
 10^{-3}$), $0.12$ ($0.15$), $0.56$ ($0.62$), and $0.99$ ($0.99$),
 respectively. 
Shown are the density field (green) overlayed with
 the ionized fraction (red/orange), and cells containing active
 sources are  shown as dots. Each slice has a thickness of
 $86\,h^{-1}\,{\rm ckpc}$, while sources shown are from a thicker 
($\sim 1.8 \,h^{-1}\,{\rm cMpc}$) region.
\label{fig:reion}}
\end{figure*}
The H-ionizing photon luminosities per halo in our cosmic reionization 
simulations are assigned in the following way. Halo catalogues are 
discrete in time, because N-body density fields are stored every 
$\sim20\,{\rm Myrs}$ and the corresponding source (halo) catalogues 
are produced at the same time. A halo of mass $M$ is assumed to have 
converted a mass $M\cdot(\Omega_{b}/\Omega_{m})\cdot f_{*}$ into stars, 
where $f_{*}$ is the star formation efficiency%
\footnote{Not all the halos convert their mass into stars. Halos inside 
H~II regions are assumed to be ``failed'' sources (i.e. their ability
to form stars is suppressed by the photoionization which created the H~II 
region) if their mass is below $10^9\,M_\odot$. More realistically,
source formation in halos of this mass range may have a gradual
dependence on halo mass: as mass decreases, it is harder to form
sources inside, and vice versa \citep{dwarf_UV}. Nevertherless, as
halo population is 
dominated by lowest-mass halos, which are most vulnerable to
photoheating in this mass range, 
we adopt our simple prescription in this
paper.}. If each 
source forms  
stars over a period of time $\Delta t$ and each stellar baryon produces 
$N_{i}$ ionizing photons per stellar lifetime and is used only once per 
$\Delta t$, and if a fraction $f_{{\rm esc}}$ of these photons escape into 
the IGM, the ionizing photon number luminosity of a halo of mass $M$ will
be given by 
\begin{equation}
Q_{i}=\frac{N_{i}\cdot f_{{\rm esc}}\cdot f_{*}\cdot M\left(\Omega_{b}/\Omega_{m}
  \right)}{\Delta t\cdot\mu m_{H}},
\label{eq:Qi}
\end{equation}
where $\mu$ is the mean molecular weight and $m_{H}$ is the mass
of a hydrogen atom. In this model, stars are produced in a burst,
and they keep radiating with fixed $Q_{i}$ for $\Delta t\simeq20\,{\rm Myrs}$.
It is noteworthy that the result does not depend on the detailed shape
of the source spectrum, but only on a frequency-integrated parameter
$N_{i}$. 

We calculate $\left\langle L_{\nu}\right\rangle $ for the LW background 
sources in a similar way. All H-ionizing sources also produce 
${\rm H}_{2}$ dissociating photons, and their 
$\left\langle L_{\nu}\right\rangle $ is also constant during the source
lifetime $\Delta t$. In each succeeding time interval, $\Delta t$, new 
sources are identified with the halo catalogue for that time-slice in 
the N-body results and are assumed to emit radiation with constant 
$\left\langle L_{\nu}\right\rangle $. This $\left\langle L_{\nu}\right\rangle$
is proportional to $(Q_i/f_{\rm esc})(N_{\rm LW}/N_i)$, where the 
proportionality constant is the dimensional factor which indicates 
the frequency-integrated number of ${\rm ergs}\,{\rm s}^{-1}\,{\rm
  Hz}^{-1}$ per LW photon  
released in the source spectrum.
We then construct the future world lines of these sources across the 
source lifetimes, $\Delta t$. At a given observing redshift, we then 
draw past light cones from every grid point in the simulation box. When 
the past light cone of an observer intersects the world line of a 
certain source, we register the comoving distance $r_{{\rm os}}$ to that 
source and its flux contribution.

We choose here a specific case from IMSP, the self-regulated reionization
case {}``f2000\_250S'' with {\em WMAP}3 background cosmology. In this
scenario, small-mass halos ($10^{8}\lesssim M/M_{\odot}\lesssim10^{9}$)
host high-efficiency emitters with top-heavy initial mass function, or
``IMF'' (e.g. massive Pop III stars). The (hydrogen-ionizing) photon 
production efficiency, $f_{\gamma}\equiv f_{*}f_{{\rm esc}}N_{{\rm i}}$, 
of these sources is approximated by $f_{\gamma}=2000$. On the other hand, 
large-mass halos ($M/{M}_{\odot}\gtrsim 10^{9}$) are assumed to host 
lower-efficiency emitters approximated by $f_{\gamma}=250$ (e.g. Pop II 
stars with Salpeter IMF). The simulation box has a volume 
$(35\, h^{-1}\,{\rm Mpc})^{3}$, with $h=0.73$. The reader is referred 
to IMSP for more details.

At this point, we need to deal with the fact that the simulation box 
($35\, h^{-1}\,{\rm cMpc}$ in this case) is smaller than the LW horizon 
($97.39\alpha\,{\rm cMpc}$). We simply use a periodic box condition, 
attaching identical boxes around the domain of calculation. We use 
$5^{3}$ identical boxes in total, choosing the central one as the 
domain of computation. One may, instead, shift and rotate boxes before 
attaching them: however, this approach would not be able to remove the 
finite-box effect completely either. In the future, we will use a much 
larger simulation box, which will naturally reduce this effect. 

We parallelized our code using the message passing interface library 
(MPI) to calculate the evolution of the spatially-varying LW background 
on distributed-memory parallel computers. We used ``Lonestar'', a 
massively-parallel supercomputer at the Texas Advanced Computing 
Center (TACC) at the University of Texas at Austin. It has a total of 
5,200 cores of dual-core Intel Xeon 5100 processors and 11.6 TB 
of aggregate memory. The particular run of LW background calculation 
presented here -- run separately from the cosmic reionization 
simulation -- took about 15 hours of computing time on Lonestar, when 
we used 256 computing cores and about 1.5 GB memory per core. The numbers 
of halos on the radiative transfer grid of $203^3$ cells in the simulation 
box of $35\,h^{-1}\,{\rm cMpc}$ were about $1.3\times 10^3$, $7\times 10^4$, 
and $1.9\times 10^5$ at $z\simeq 15$, 10, and 8, respectively. Note that 
the effective total number of sources for our LW background calculation 
is about 125 times the number of sources in the simulation box of this 
particular size, due to the length scale of $r_{\rm  LW}$. 

\subsection{Evolution of the Globally-Averaged Ionizing and Dissociating 
Radiation Backgrounds}
\label{sub:average}

\begin{figure*}[ht]
\includegraphics[width=0.5\textwidth]{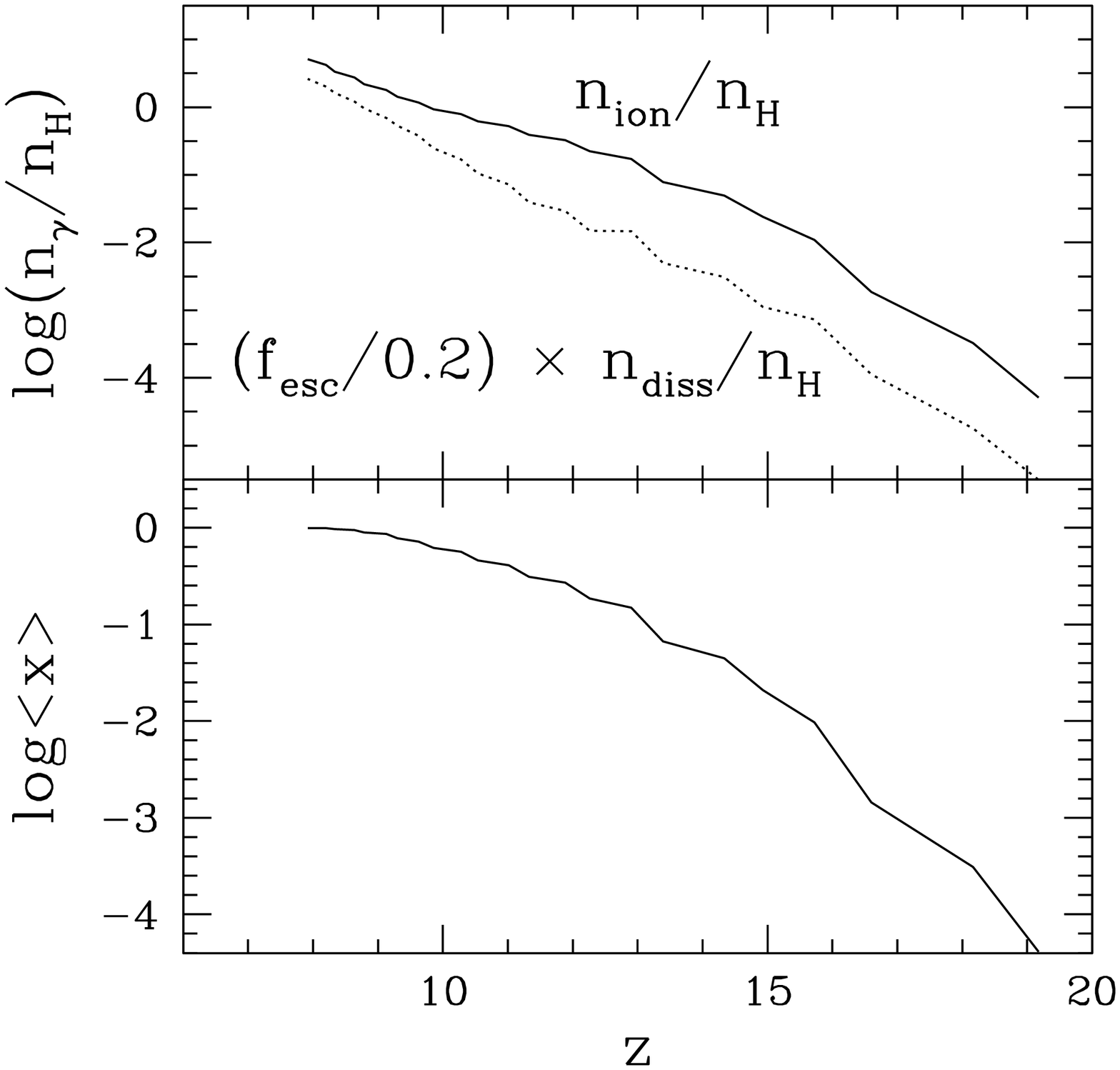}
\includegraphics[width=0.5\textwidth]{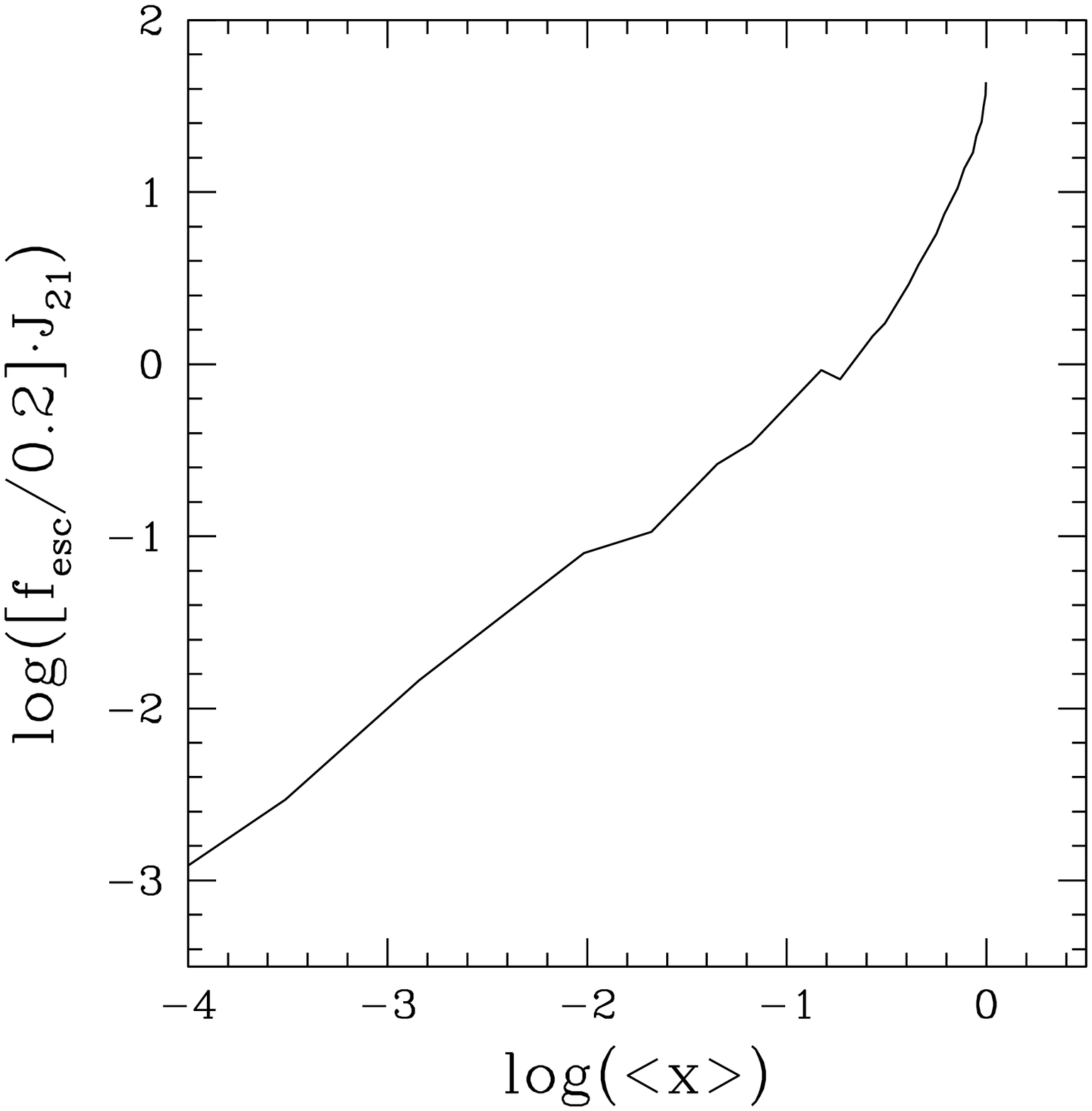}

\caption{
Global history of cosmic reionization and the dissociation background:
(Left panel)(top) Evolution of the photon to baryon ratios, where 
$n_{{\rm ion}}$ is the total ionizing photon number density accumulated 
until given redshift $z$, while $n_{{\rm LW}}$ is the {}``instantaneous'' 
dissociating photon number density at $z$ given by 
$n_{{\rm LW}}\equiv\frac{4\pi}{c}\int_{11.5\,{\rm eV}}^{13.6\,{\rm eV}}
\frac{J_{\nu}}{h\nu}d\nu\simeq\frac{4\pi}{ch}\frac{2.1\,{\rm eV}}
{12.6\,{\rm eV}}\langle J_{\nu}\rangle$.
(bottom): Global mean ionized fraction $\langle x\rangle$ vs. redshift. 
(Right panel): Global mean dissociating intensity, 
${\langle J}_{{\rm LW},\,21}\rangle$, multiplied by $(f_{{\rm esc}}/0.2)$, vs.
the global mean ionized fraction.
\label{fig:nn}}
\end{figure*}

The growth and geometry of the ionized fraction of the universe during 
the EOR is illustrated by the selected time-slices shown in 
Figure~\ref{fig:reion}. The corresponding evolution of the 
globally-averaged ionized fraction and the ionizing and dissociating 
radiation backgrounds is summarized in Figure \ref{fig:nn}, with several 
interesting features revealed. First, at $z\gtrsim10$, small-mass halos 
dominate the large-mass halos in contributing both ionizing and 
dissociating photons. This is easily understood in the framework of the
standard $\Lambda$CDM cosmology, because the population of low-mass 
halos dominates over that of high-mass halos, both in numbers and in 
total mass. These low-mass sources, however, become 
``self-regulated'' as the universe gets more ionized (since they are 
suppressed as sources in the H~II regions) and are later almost fully 
suppressed at $z\lesssim10$, while the collapsed fraction in the 
unsuppressible (higher-mass) sources only grows with time. Thus, both 
reionization and dissociation are dominated by high-mass halos at 
$z\lesssim10$. Second, $n_{{\rm i}}/n_{{\rm LW}}$ becomes smaller at 
$z\lesssim10$ than at $z\gtrsim10$. This is simply due to the transition 
of major source type from Pop III to Pop II, because 
$\left(N_{{\rm i}}/N_{{\rm LW}}\right)_{{\rm III}}>
       \left(N_{{\rm i}}/N_{{\rm LW}}\right)_{{\rm II}}$. 

We note that the reionization history depends on the adopted values
of the efficiency parameters, $f_{\gamma}$, for low-mass and high-mass 
halos. The ionizing background is degenerate in $f_{*}$, $f_{{\rm esc}}$
and $N_{{\rm i}}$, in fact, as long as their product $f_{\gamma}$ is fixed.
This is not true for the dissociating background, however, which breaks 
the the degeneracy between $f_{{\rm esc}}$ and $N_{{\rm i}}$, in general. 
Because dissociating photons --- with emitted energy ranging from 
$11.2\,{\rm eV}$ to $13.6\,{\rm eV}$ --- are largely unattenuated by their 
own interstellar medium inside the source halos, their escape fraction 
from the source halo is essentially unity. Therefore, for a given 
$f_{\gamma}$, or a given reionization history,
\[
J_{{\rm LW},\,21}\propto1/f_{{\rm esc}}.
\]
Hardness of the spectral energy distribution (SED) of the source,
which can be characterized by $N_{{\rm i}}/N_{{\rm LW}}$, also
affects $J_{{\rm LW},\,21}$. For a given $f_{\gamma}$, 
\[
J_{{\rm LW},\,21}\propto\left(N_{{\rm i}}/N_{{\rm LW}}\right)^{-1}.
\]
Assuming a top-heavy IMF, Pop III objects have 
$\left(N_{{\rm i}}/N_{{\rm LW}}\right)_{{\rm III}}\approx15$
\citep[e.g.][]{2000ApJ...528L..65T,2001ApJ...552..464B}.
Pop II objects with a Salpeter IMF, on the other hand, have 
$\left(N_{{\rm i}}/N_{{\rm LW}}\right)_{{\rm II}}\approx1$
(\citealt{2000ApJ...528L..65T} and references therein).

The spatially-averaged mean intensity is already as high as 
$\langle J_{{\rm LW},\,21}\rangle=0.1$ by $z\approx15$, when the mean 
ionized fraction of the universe is only $\langle x\rangle\approx0.02$, 
if the fiducial value of 
$f_{{\rm esc,\, III}}=0.2$ is used for Pop III objects. $J_{{\rm LW},\,21}$ 
is not affected by $f_{{\rm esc,\, II}}$ until $z\approx15$, because Pop~II 
objects start to emerge only after $z\approx15$. As described above, 
$J_{{\rm LW},\,21}$ is proportional to $1/f_{{\rm esc}}$. 
For instance, if $f_{{\rm esc,\, III}}=1$ instead of 0.2, then 
$\langle J_{{\rm LW},\,21}\rangle=0.1$ will be reached later after 
$z\approx15$. $\langle J_{{\rm LW},\,21}\rangle$ depends on both 
$f_{\rm esc,\, III}$ and $f_{{\rm esc,\, II}}$ at $z\lesssim15$, and if 
$f_{{\rm esc,\, II}}$=1 as well, then 
$\langle J_{{\rm LW},\,21}\rangle=0.1$ at $z\approx13$, or when 
$\langle x\rangle\approx0.1$.

\begin{figure}[ht]
\includegraphics[width=0.5\textwidth]{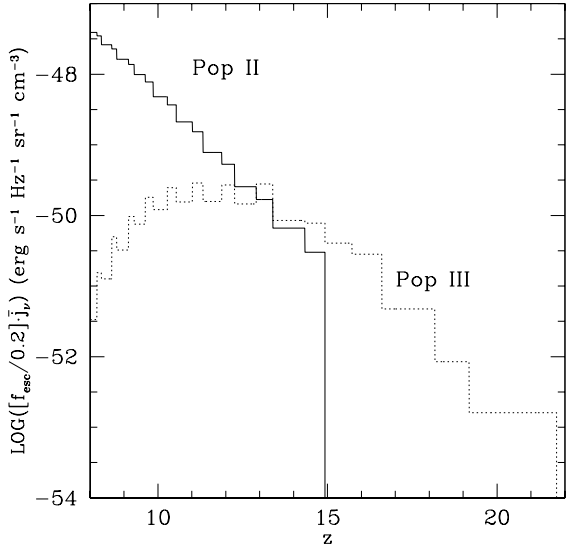}
\caption{Comoving emission coefficients $\bar{j}_{\nu_{{\rm s}}}(z_{{\rm s}})$
from Pop III objects (dotted) residing in low-mass ($10^{8}\le
M/{ M}_{\odot}\le10^{9}$) halos and Pop II objects (solid) residing 
in high-mass ($M/{M}_{\odot}\ge10^{9}$) halos. These are constant 
during the source lifetime $\Delta t\approx20\,$Myrs, because halo 
collapsed fractions, obtained from the self-regulated simulation 
results (IMSP), are also assumed to be constant during $\Delta t$, 
which is also identical to the time interval between adjacent N-body
outputs.\label{fig:jnu}}
\end{figure}

\begin{figure}[ht]
\includegraphics[width=0.5\textwidth]{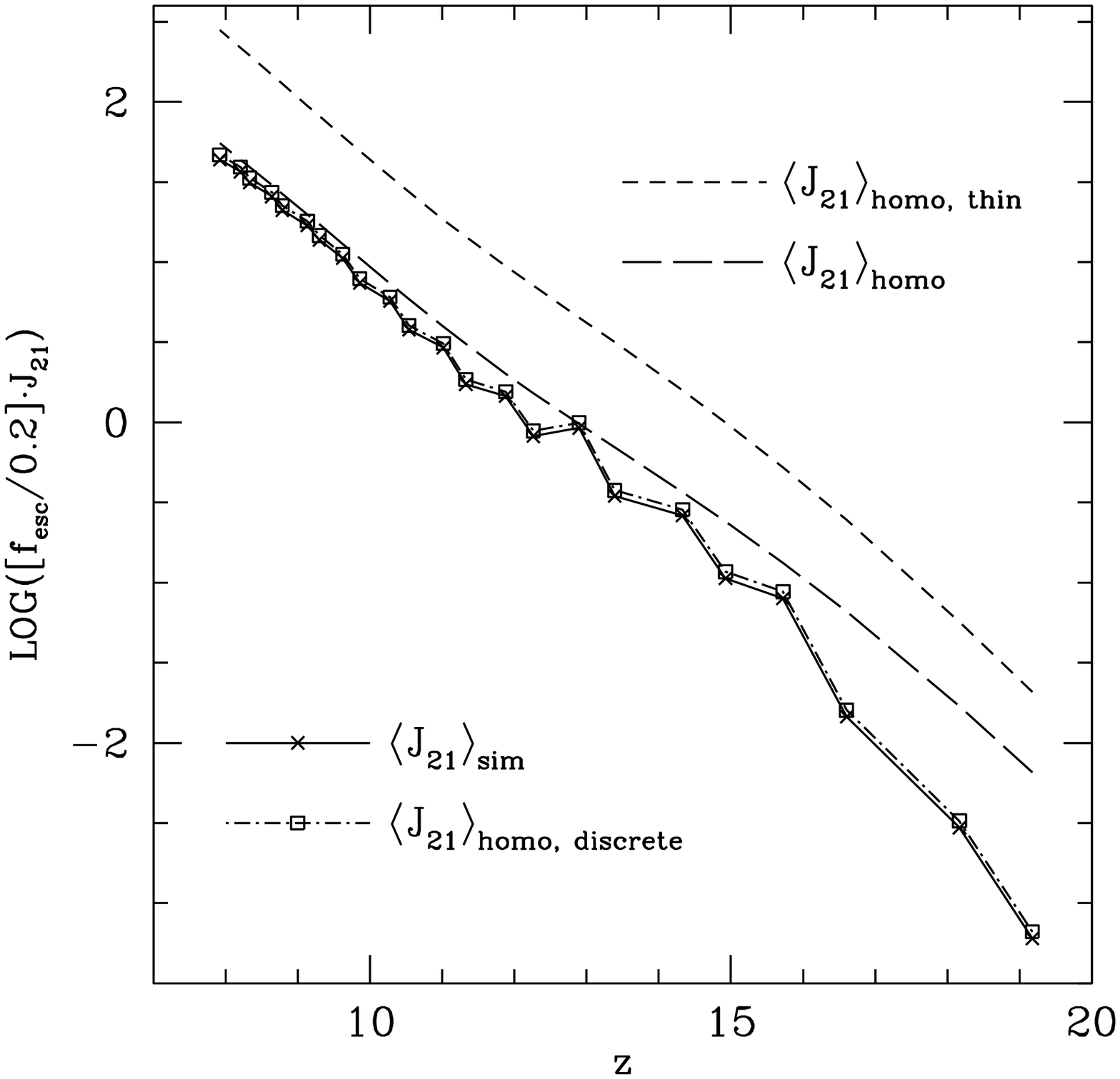}
\caption{Evolution of the global average LW background intensity 
(in units of $10^{-21}\rm\,erg\,cm^{-1}\,s^{-1}\,Hz^{-1}\,ster^{-1}$)
$\langle J_\nu\rangle_{\rm LW,21}$: $\langle J_{\nu}\rangle_{\rm LW,21}$ 
(solid-cross), the intensity obtained by averaging the LW background 
field over the simulation volume at each redshift $z$, 
$\langle J_{\nu}\rangle_{\rm LW, 21, hom}$ (dot-dashed-square),
the mean intensity in a homogeneous universe but with the
emission coefficient depicted in Figure~\ref{fig:jnu} (i.e. distrete 
in time). We also plot the average intensity, 
$\langle J_\nu\rangle_{LW, 21, \,{\rm semi}}$ (long-dash), based 
upon our smooth fitting formulae (i.e. continuous in time) for the 
collapsed fractions of both low-mass ($10^{8}\le M/{ M}_{\odot}\le10^{9}$) 
and high-mass ($M/{M}_{\odot}\ge 10^{9}$) halos. For comparison, 
$\langle J_\nu\rangle_{LW, 21, \,{\rm semi},\,{\rm OT}}$ (short dash) is the 
intensity when we neglect the optical depth to H Lyman lines but take 
the horizon as the distance over which Lyman limit photons (13.6 eV) 
redshift to the minimum energy (11.5 eV) photons of our interest.
\label{fig:J21comp}}
\end{figure}

It is useful to compare our numerical results for the space-averaged LW
intensity with the homogeneous universe approximation. The global mean 
dissociating intensity $\langle J_{{\rm LW},\,21}\rangle$ is easily 
predictable if the fraction of mass collapsed into stars and the number
of dissociating photons created per stellar baryon are known. Consider
a universe where sources are homogeneously distributed, with a collapsed 
fraction which is identical to that obtained numerically from our N-body 
simulation. The emission coefficient (in 
${\rm erg}\,{\rm s}^{-1}\,{\rm Hz}^{-1}\,{\rm sr}^{-1}\,{\rm cm}^{-3}$)
is then only a function of redshift as follows:
\begin{equation}
j_{\nu_{{\rm s}}}(z_{{\rm s}})
  =\frac{1}{4\pi}\epsilon_{\nu_{{\rm s}}}\rho_{m}(z_{{\rm s}})
  \frac{\Omega_{b}}{\Omega_{m}}f_{{\rm coll}}(z_{{\rm s}})f_{*},
\label{eq:jnuprop}
\end{equation}
where 
$\epsilon_{\nu_{{\rm s}}}({\rm erg}\,{\rm s}^{-1}\,{\rm Hz}^{-1}\,{\rm g}^{-1})$
is the emissivity at emitted frequency $\nu_{{\rm s}}$, $\rho_{m}(z_{{\rm s}})$
is the mean mass density at redshift $z_{{\rm s}}$, and 
$f_{{\rm coll}}(z_{{\rm s}})$ is the source halo collapsed fraction. Note that 
this is the proper emission coefficient: the comoving emission coefficient is 
given by
\begin{eqnarray}
\bar{j}_{\nu_{{\rm s}}}(z_{{\rm s}}) 
  &\equiv& j_{\nu_{{\rm s}}}(z_{{\rm s}})/(1+z_{{\rm s}})^{3}\nonumber \\
  & = & \frac{1}{4\pi}\epsilon_{\nu_{{\rm s}}}\rho_{m,\,0}
  \frac{\Omega_{b}}{\Omega_{m}}f_{{\rm coll}}(z_{{\rm s}}),
\label{eq:jnucom}
\end{eqnarray}
where $\rho_{m,\,0}$ is the mean mass density at present. Finally,
the (proper) mean intensity in this homogeneous universe is obtained
by
\begin{equation}
\langle J_{\nu}\rangle_{\rm homo}(z_{{\rm obs}})
   =(1+z_{{\rm obs}})^{3}\int_{0}^{r_{{\rm LW}}}
   \frac{dr_{{\rm os}}}{1+z_{{\rm s}}}\,\bar{j}_{\nu_{{\rm s}}}(z_{{\rm s}})
   \cdot f_{{\rm mod}}(r_{{\rm os}}),
\label{eq:Jnuhom}
\end{equation}
where the source redshift $z_{{\rm s}}$ is implicitly related to
$r_{{\rm os}}$ by equation (\ref{eq:r_os2}). The emission coefficient 
$\bar{j}_{\nu_{{\rm s}}}(z_{{\rm s}})$ is shown in Figure~\ref{fig:jnu}, 
where the ``stair-steps'' reflect the fact that, as described in 
\S~\ref{sub:summing} , we use halo catalogues that are discrete in time, 
which results in a discontinuous evolution of $f_{{\rm coll}}$ as well. 
The resulting $\langle J_{\nu}\rangle_{\rm LW,\,homo,\,discrete}$ plotted in 
Figure~\ref{fig:J21comp} agrees well with the globally-averaged value
of $\langle J_{\nu}\rangle_{\rm LW,\,sim}$ from our simulations, as expected.
For comparison, we also plot in Figure~\ref{fig:J21comp} the 
homogeneous approximation result when the discrete, time-stepped 
collapsed fractions, $f_{\rm coll}$, shown in Figure~\ref{fig:jnu} are
replaced by a smoothly varying mass function based on fitting the 
N-body results over time. The importance of the H Lyman line opacity
in attenuating the LW photons is illustrated by the quantity 
$\langle J\rangle_{\rm LW,\,homo,\,thin}$, also plotted in 
Figure~\ref{fig:J21comp}, the unattenuated mean intensity in the 
homogeneous approximation if we neglect H Lyman line opacity but take 
account of the ultimate horizon which corresponds to the distance from 
which Lyman limit photons at 13.6 eV redshift to the minimum energy 
(11.5 eV) of interest here for the LW dissociation rate.


\begin{figure}[ht]
\includegraphics[width=0.5\textwidth]{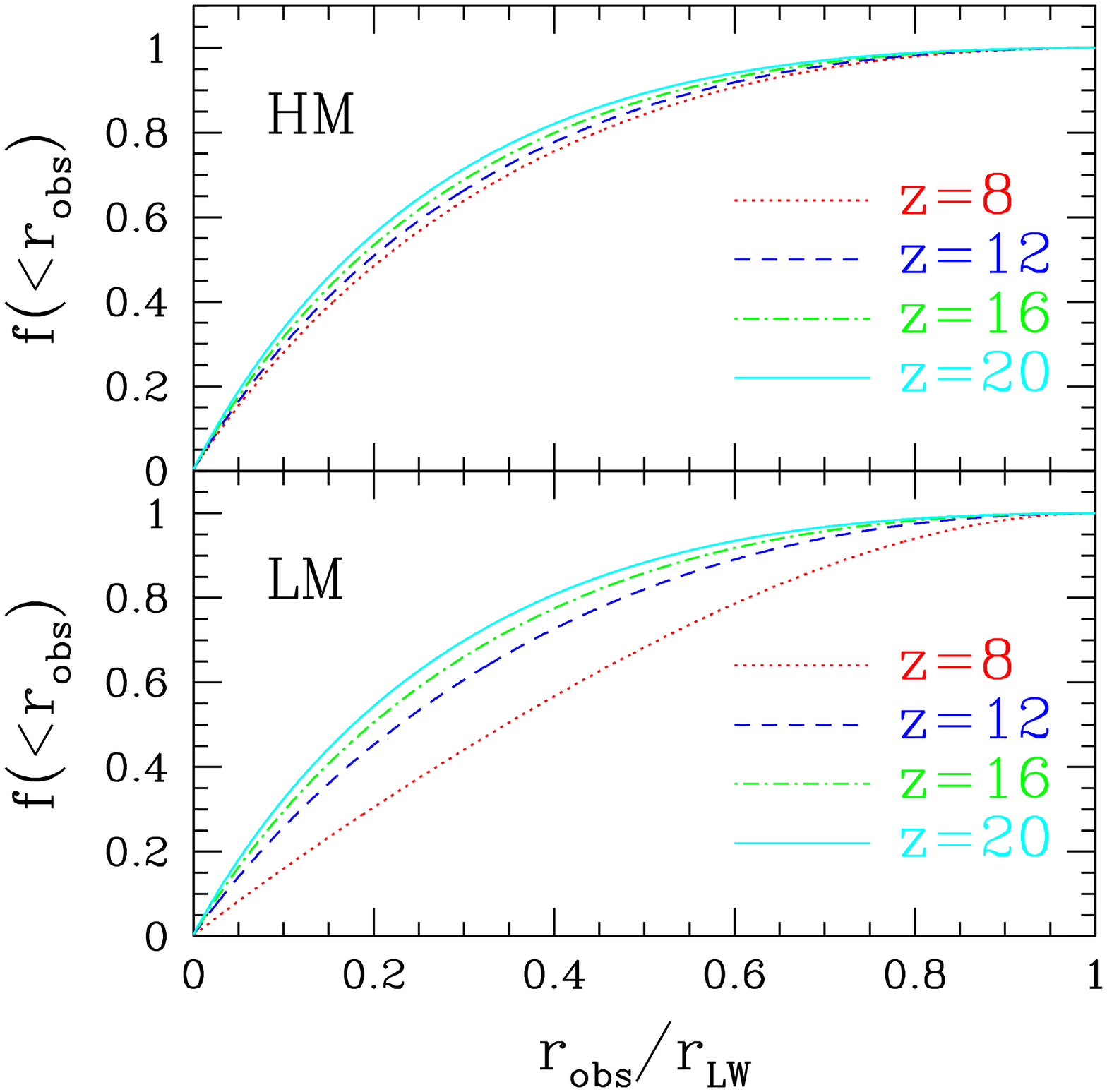}
\includegraphics[width=0.5\textwidth]{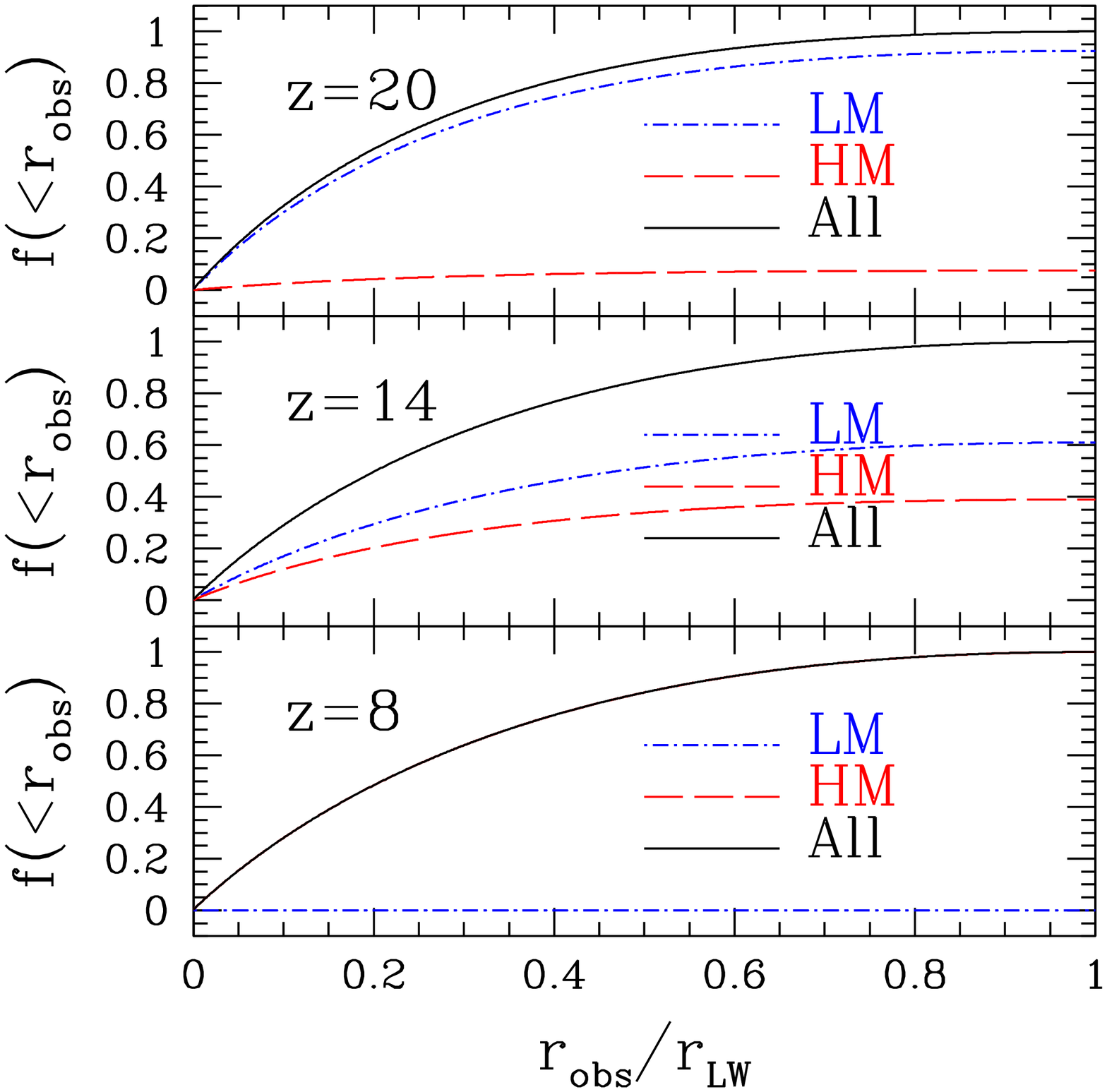}
\caption{Fractional contribution to $J_{\nu}$ from sources inside a sphere
of comoving radius $r_{{\rm os}}$ (normalized to $r_{{\rm LW}}$) from the 
observer. This is separated into two categories by mass: ({\em top-left}) 
contribution to $J_{{\rm LW},\,{\rm HM}}$, or the intensity $J_{\rm LW}$ only 
by sources in high-mass ($M/{M}_{\odot}\ge 10^{9}$) halos. ({\em
  bottom-left}) contribution to  
$J_{{\rm LW},\,{\rm LM}}$, or the intensity $J_{\rm LW}$ only by sources in 
low-mass ($10^{8}\le M/{ M}_{\odot}\le10^{9}$) halos. 
({\em right}) Fractional contribution to the toal $J_{\rm LW}$ by low-mass halo
sources (LM, dot-dashed), high-mass halo sources (HM, dashed), and all
sources (All, solid). Low-mass halo contribution, dominating $J_{\rm LW}$
at $z\sim 20$, becomes comparable to high-mass halo contribution at
$z\sim 14$, and becomes negligible at $z\sim
8$. Considering both (All) contributions, about 80\% of $J_{\rm LW}$ comes
from $r \lesssim [0.35-0.45] r_{\rm LW}$ at all redshifts. This
indicates that both
isotropic and fluctuating components of the LW background are
dominated by nearby sources at all times.
\label{fig:fignorm}
}
\end{figure}

Which sources are the dominant contributors to $J_{\nu}$, sources 
{\it near} to or sources {\it far} from the observer? Let us 
consider the same homogeneous universe as described above. If photons 
were not attenuated ($f_{{\rm mod}}=1$),
and the comoving emission coefficient $\bar{j}_{\nu_{{\rm s}}}(z_{{\rm s}})$
remained constant over time, then spherical shells with identical proper 
thickness $\Delta\left(\frac{dr_{{\rm os}}}{1+z_{{\rm s}}}\right)$ would 
contribute equally to $J_{\nu}$, as seen in equation (see \ref{eq:Jnuhom}). 
This is not the case, however. The factor $f_{{\rm mod}}$ increases as 
$r_{{\rm os}}$ decreases (Figure \ref{fig:fmod}), and $f_{{\rm coll}}$ 
usually increases as $z_{{\rm s}}$ decreases as well. These two factors 
combine to make nearby sources more important. An opposite trend 
can occur, however, if we consider only the sources inside low-mass 
halos, because the evolution of their emission coefficient is not 
monotonic, as seen in Figure~\ref{fig:jnu}. That nonmonotonicity is not 
enough to completely offset the increase of $f_{{\rm mod}}$, however. 
Defining the fractional contribution from spheres of varying radius 
$r_{{\rm os}}$ as
\begin{eqnarray}
f(<r_{{\rm os}})&\equiv&\int_{0}^{r_{{\rm os}}}
\left(\frac{dr_{{\rm os}}}{1+z_{s}}\right)\,\bar{j}_{\nu_{{\rm s}}}(z_{{\rm
    s}})\cdot f_{{\rm mod}}(r_{{\rm os}})
\nonumber \\
&&/\int_{0}^{r_{{\rm LW}}}\left(\frac{dr_{{\rm os}}}{1+z_{s}}\right)\,
\bar{j}_{\nu_{{\rm s}}}(z_{{\rm s}})\cdot f_{{\rm mod}}(r_{{\rm os}}),
\end{eqnarray}
we have calculated $f(<r_{{\rm os}})$ both for low-mass halo sources
and high-mass halo sources. The results for the low-mass (suppressible)
and high-mass (not suppressible) halos are plotted in Figure~\ref{fig:fignorm}.
In the case of $J_{{\rm LW},\,{\rm HM}}$, for example, while all sources 
within $r_{\rm LW}$ can contribute, $\sim 80 \%$ of $J_{{\rm LW},\,{\rm HM}}$ 
is contributed by sources at $r\lesssim [0.35 - 0.45]r_{\rm LW}$. 
In the case of $J_{{\rm LW},\,{\rm LM}}$, the overall trend is similar to
that for $J_{{\rm LW},\,{\rm HM}}$, except at $z\simeq 8$. Even at this redshift,
however, an observer primarily needs to look back only to 
$r \simeq 0.4 r_{\rm LW}$, since the high-mass halos dominate the total 
emissivity at late times, as seen in Figures~\ref{fig:jnu} and
\ref{fig:fignorm}, and dominate  
$J_{\rm LW}$, as well, therefore.  

While our comparison of the simulated, globally-averaged LW background 
with that from the homogeneous approximation shows that the latter is
good as long as we give it the correct, self-regulated, space-averaged
mass function of source halos, only the {\it simulations} can tell us 
about the spatial {\it variations} in the LW background and their 
evolution, as well. We will show in the following section that a 
significant spatial fluctuation of the LW background does arise during 
the epoch of reionization. This, in principle, could induce a 
fluctuating feedback effect on the star formation in minihalos, thereby
altering the apparent pattern of their clustering from that which arises 
gravitationally due to structure formation.

Note that the possible effect by the finite size of the box, which is smaller
than the LW horizon, does not seem to affect our quantitative
conclusion too much. First, as described above, about 80\% of the LW
intensity comes from $r\lesssim[0.35-0.45] r_{\rm LW}$, which is
just about the 
box size used. This implies that the flucuation of LW intensity as well
as its mean is contirubted mostly by nearby sources. Second,
according to a suite of structure formation simulations we have
performed, halo mass functions dn/dM from these simulations do not
show too much suppression at a mass scale corresponding to the mass
of the box, where most of the suppression of power is expected to
occur. Mass functions from these simulations (in varying dynamical 
ranges), accordingly, connect smoothly when plotted on a single
viewgraph. Because LW intensity originates from sources of
reionization residing in cosmological halos, power of the LW
intensity fluctuation at the box scale would not be suppressed just
as the mass function dn/dM at the box scale is not.

\subsection{Spatial Fluctuations of the LW background}

Until now, the possible spatial fluctuations of the LW background have
been neglected, due partly to the fact that $r_{{\rm LW}}$ is a 
cosmologically large scale. One might naively have expected, therefore, 
that the LW intensity fluctuations inside the LW band photon horizon, or 
$r_{{\rm LW}}$, are negligibly small, since the fluctuations in the 
space-density of matter are small when averaged on such large scales at such
an early epoch. Nevertheless, as we will show, we find huge fluctuations
in the LW radiation field. How is this possible? In this section, we 
present our results for the fluctuations in the LW radiation field and 
explain their origin.

\begin{figure*}[ht]
\includegraphics[width=.4\textwidth]{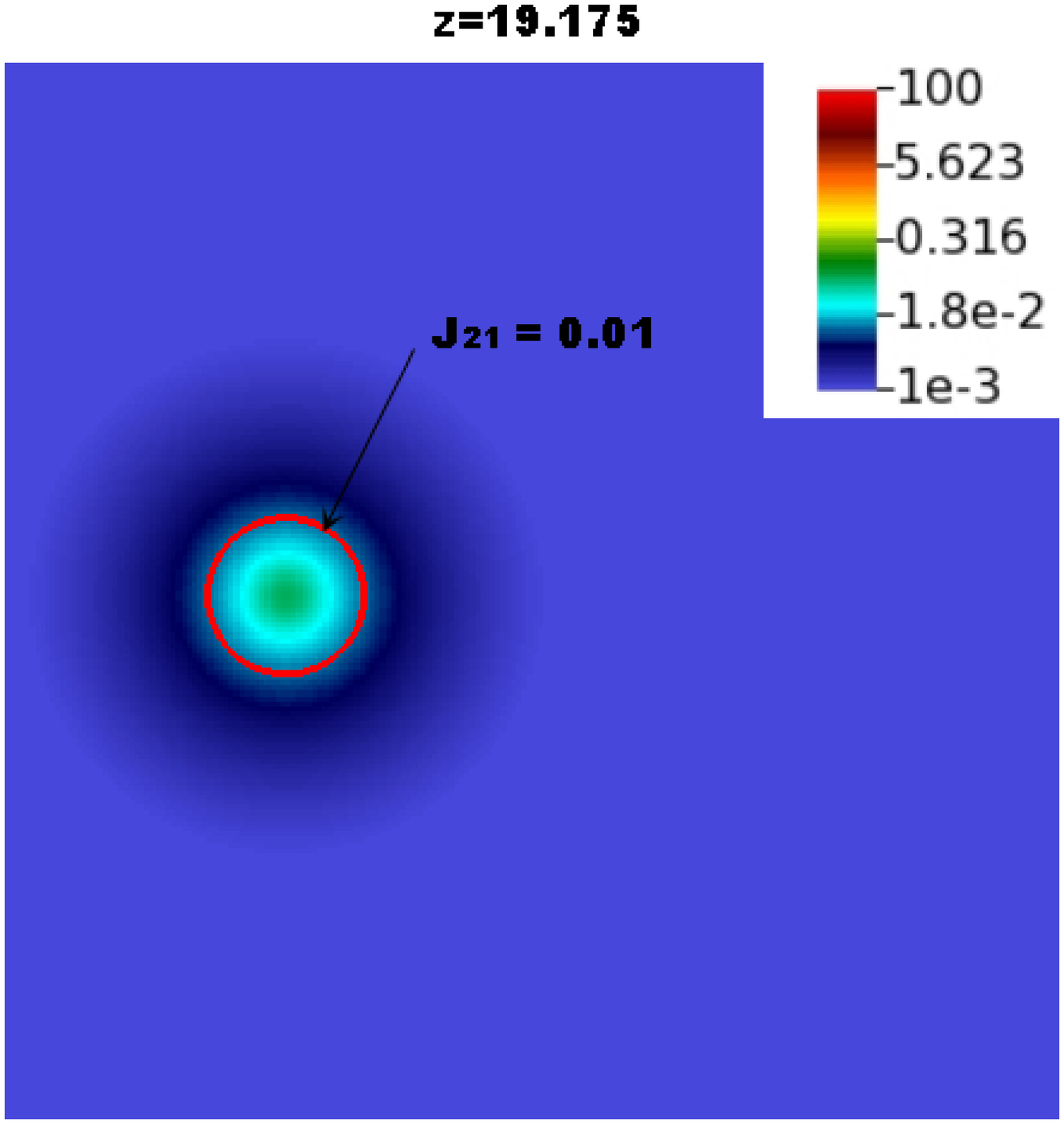}
\includegraphics[width=.4\textwidth]{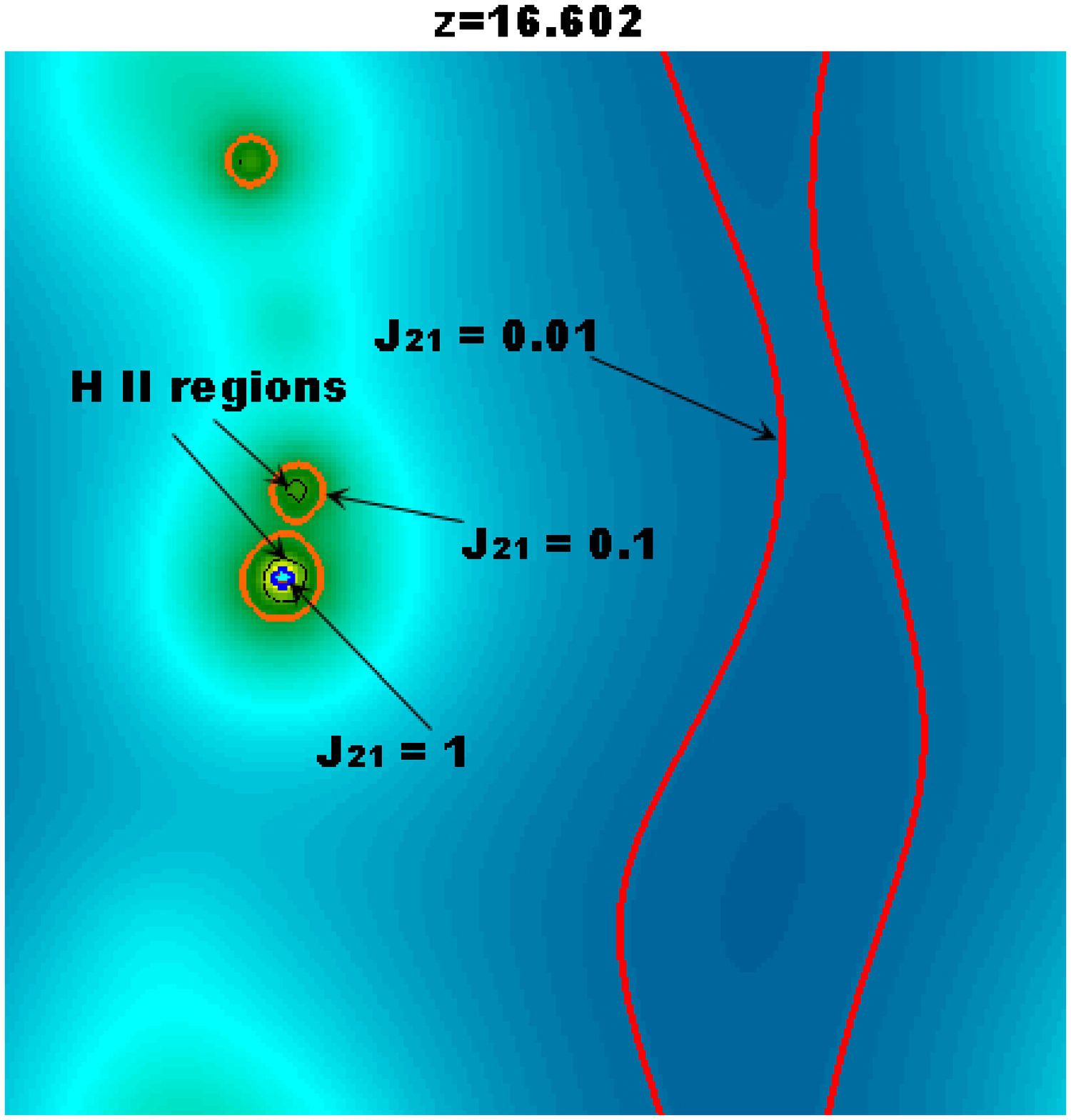}

\includegraphics[width=.4\textwidth]{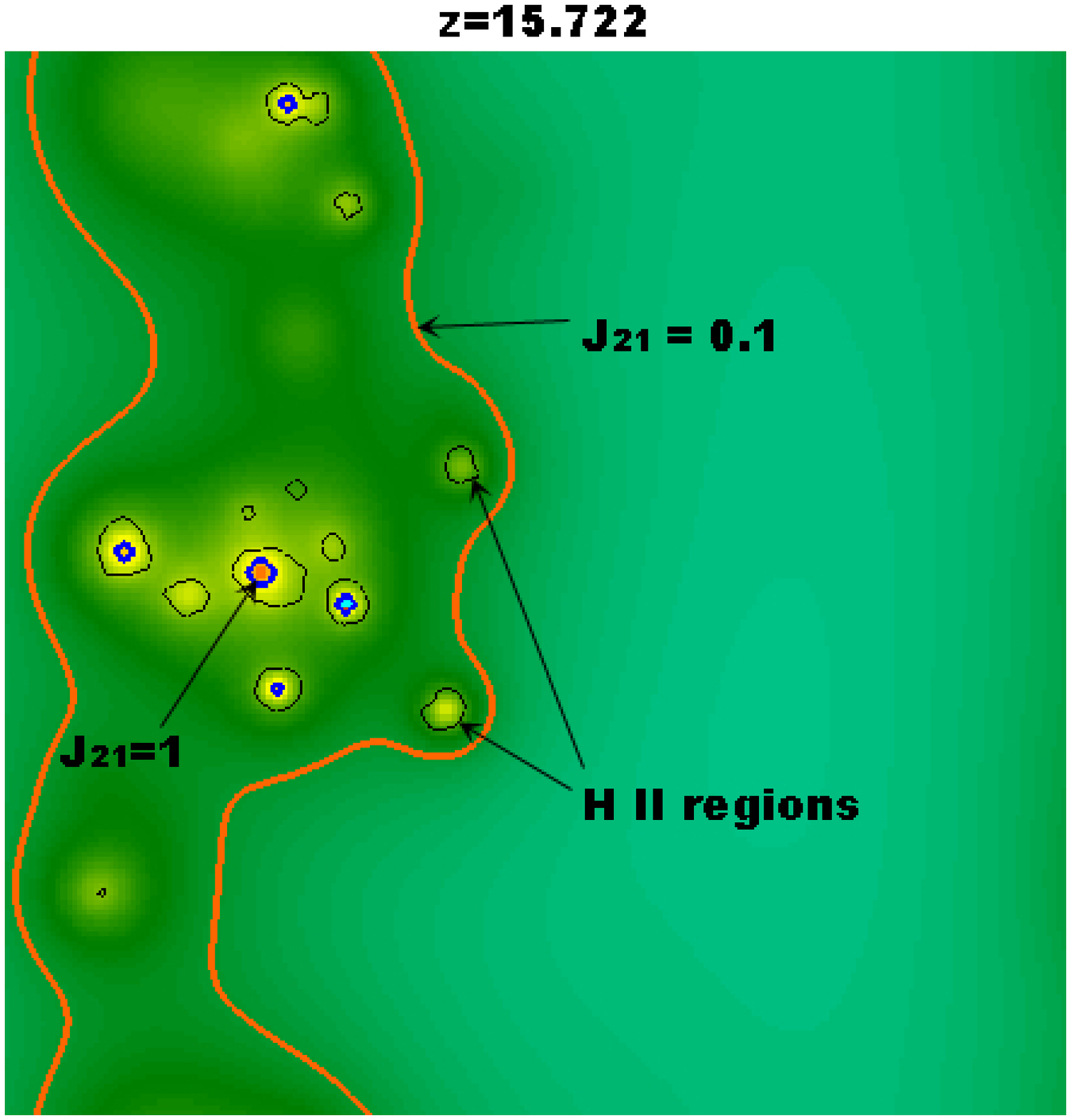}
\includegraphics[width=.4\textwidth]{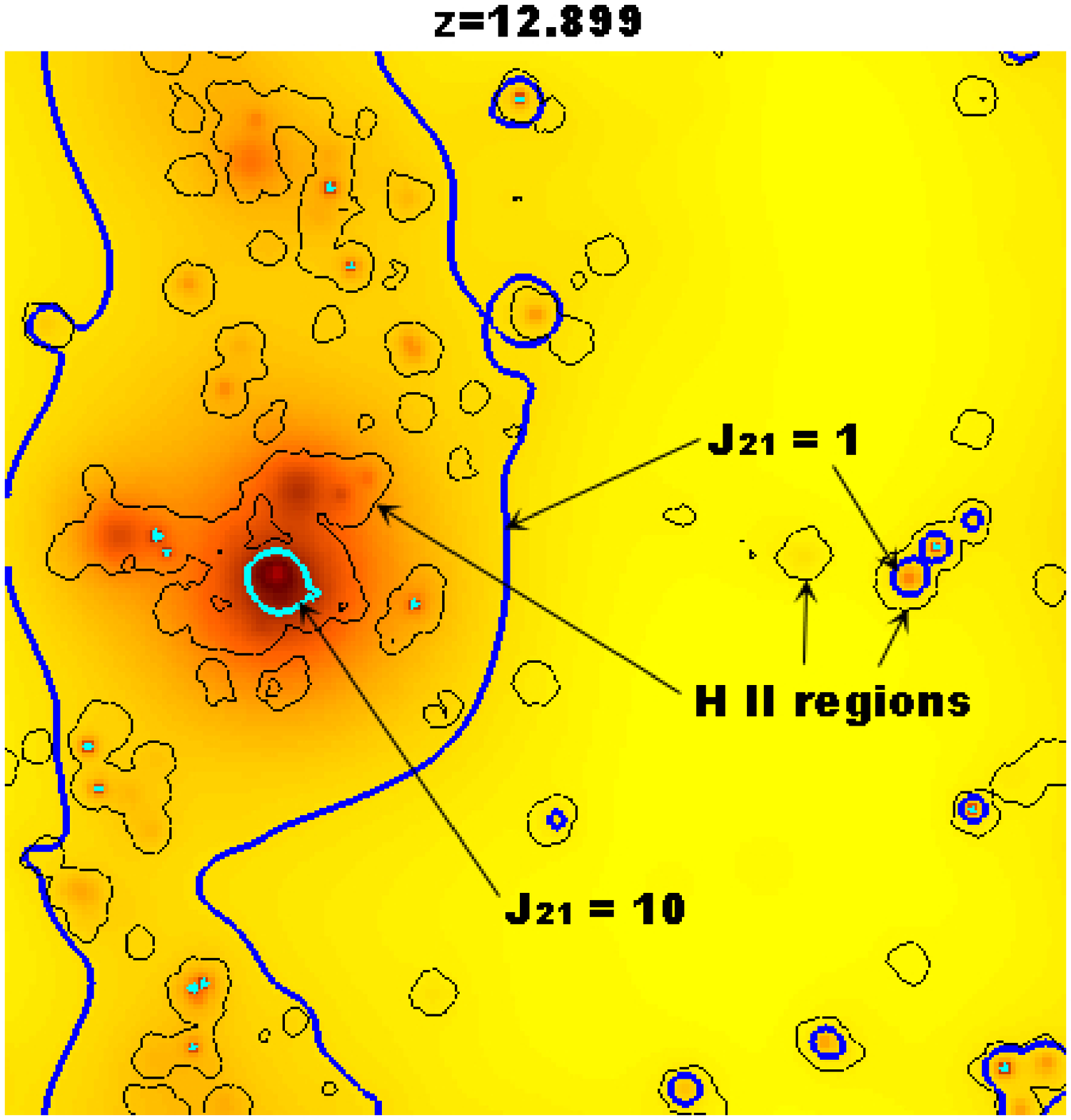}

\includegraphics[width=.4\textwidth]{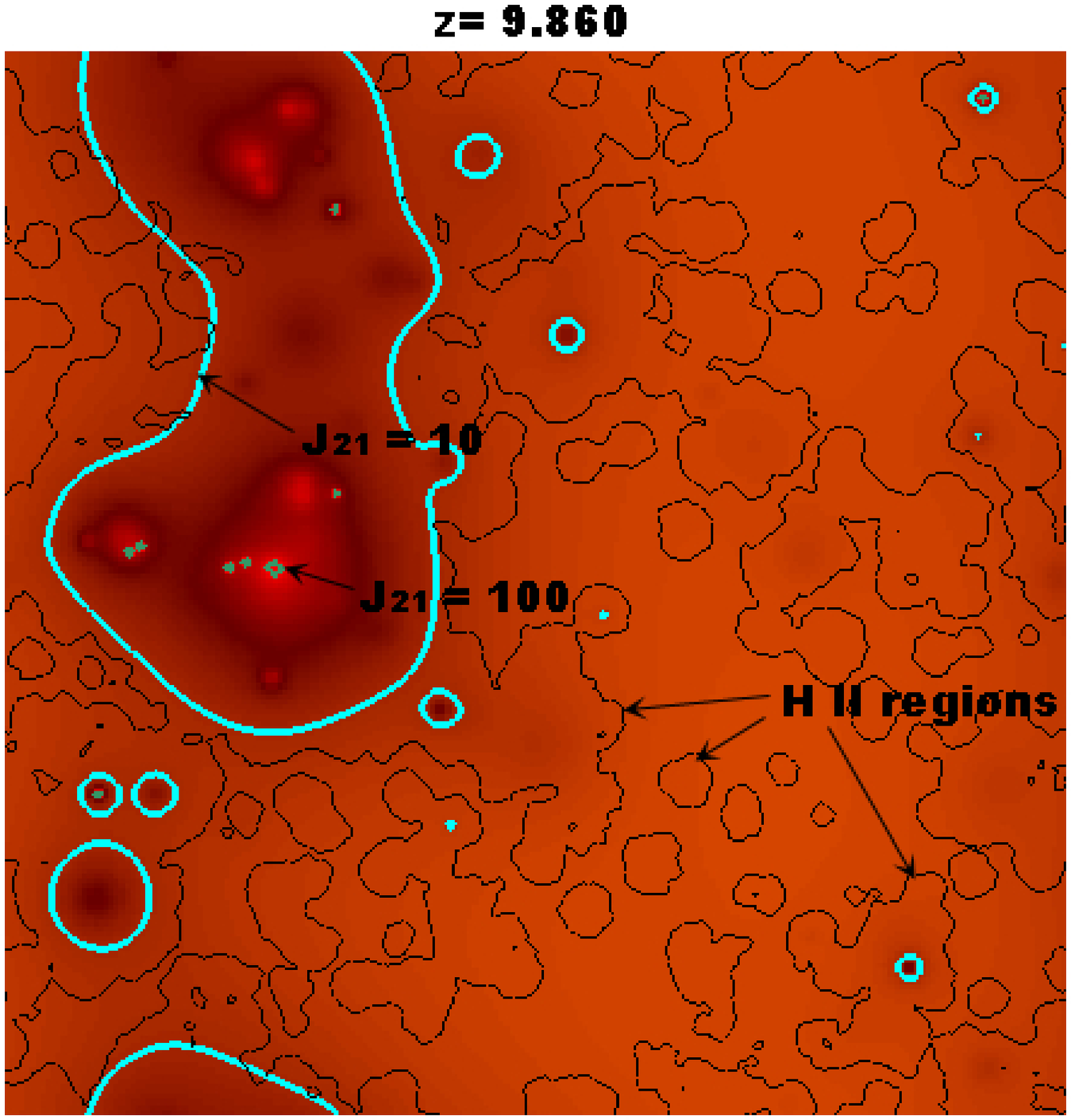}
\includegraphics[width=.4\textwidth]{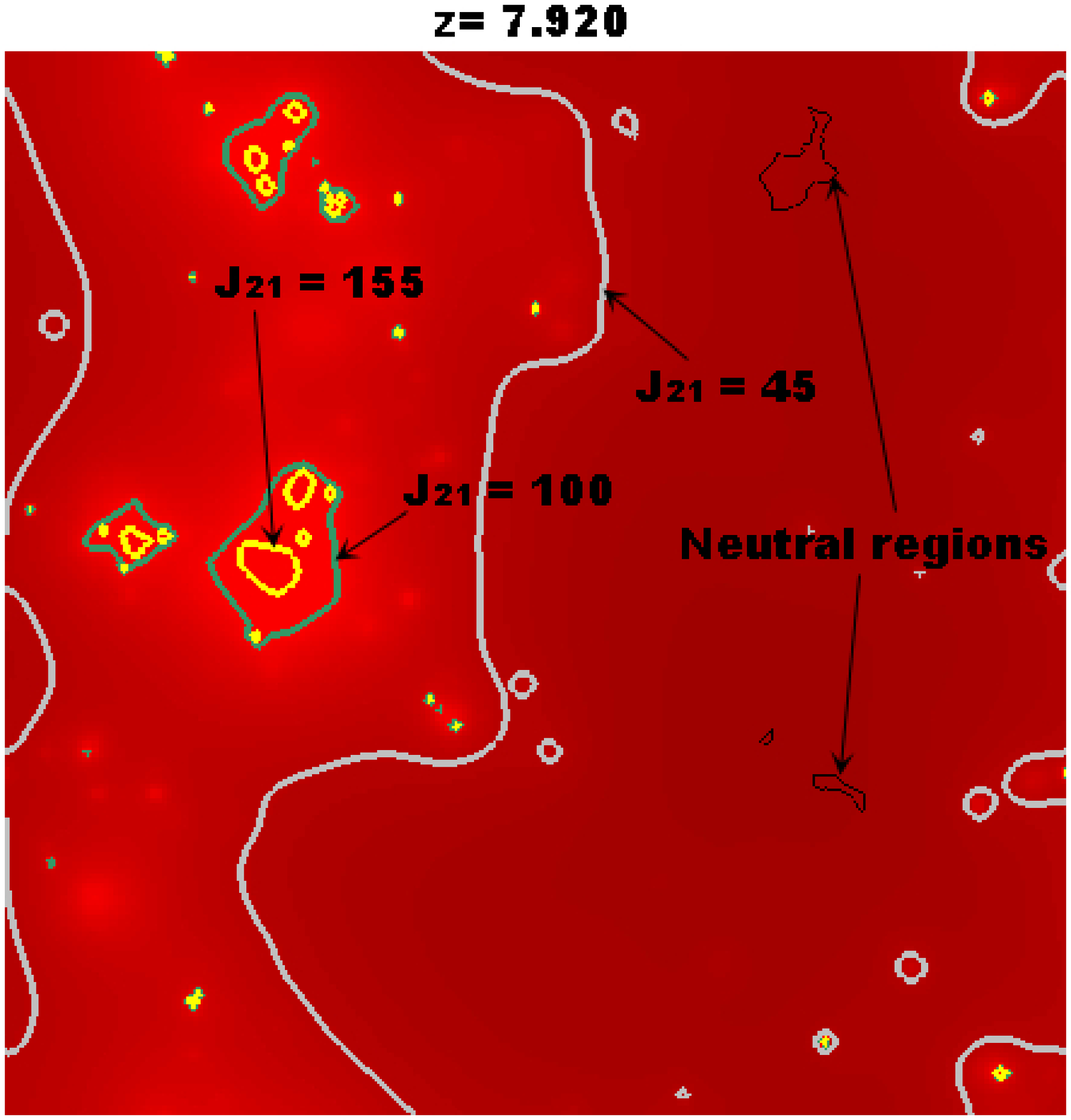}

\caption{Simulation spatial images showing the isocontours of patchy 
reionization and the patchy ${\rm H}_{2}$ dissociating background on
a planar slice through the box of volume $(35\, h^{-1}\,{\rm cMpc})^{3}$
at different epochs. The level of $J_{{\rm LW},\,21}$ on the grid is 
depicted by different colors, with the range $[10^{-3} - 10^2]$, shown 
on the inset of the top-left panel. On top of each $J_{{\rm LW},\,21}$
color-map, contours of thick colored lines represent different 
$J_{{\rm LW},\,21}$ levels (red, orange, blue, cyan, and green 
corresponding to $J_{{\rm LW},\,21}=$0.01, 0.1, 1, 10, and 100, 
respectively). The black lines represent the ionization fronts 
characterized by $x=0.5$.
\label{fig:J21cuts}}
\end{figure*}

In Figure~\ref{fig:J21cuts}, we show the $J_{{\rm LW},\,21}$ field on a 
planar slice (with fixed comoving coordinates) inside the simulation box 
at different redshifts. $J_{{\rm LW},\,21}$ varies significantly in space
at all redshifts. For instance, at $z=16.602$, the overall variance of 
$J_{{\rm LW},\,21}$ is about two orders of magnitude. Figure~\ref{fig:J21cuts} 
shows that as many as 3 contour levels are observed simultaneously in
the same image plane at this redshift, corresponding to $J_{{\rm LW},\,21}=0.01$,
0.1, and 1, respectively. This example clearly debunks any previous
assumption that might have been made that only small fluctuations would
exist in the LW background inside $r_{{\rm LW}}$.

\begin{figure*}[ht]
\includegraphics[width=0.5\textwidth]{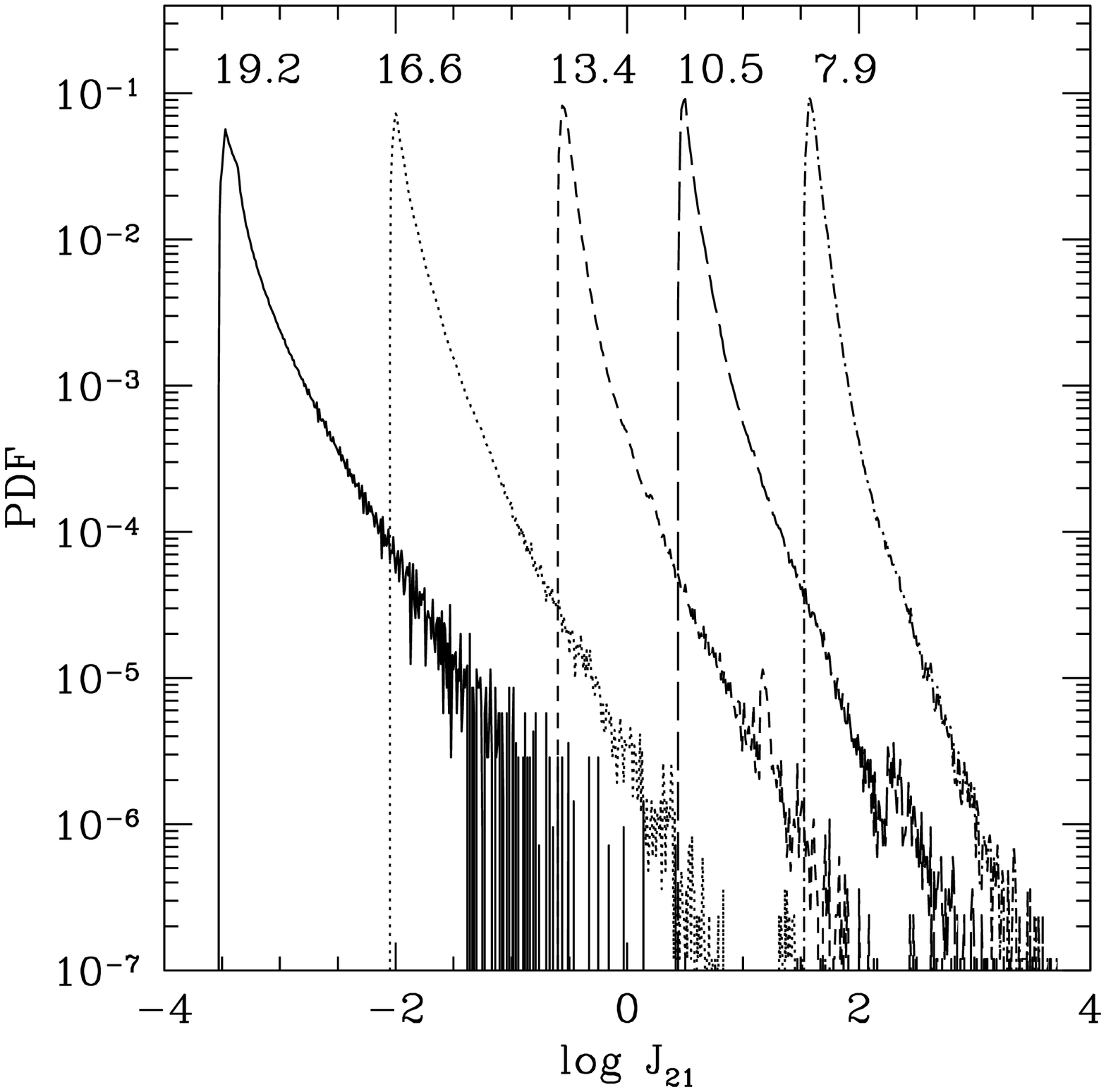}
\includegraphics[width=0.5\textwidth]{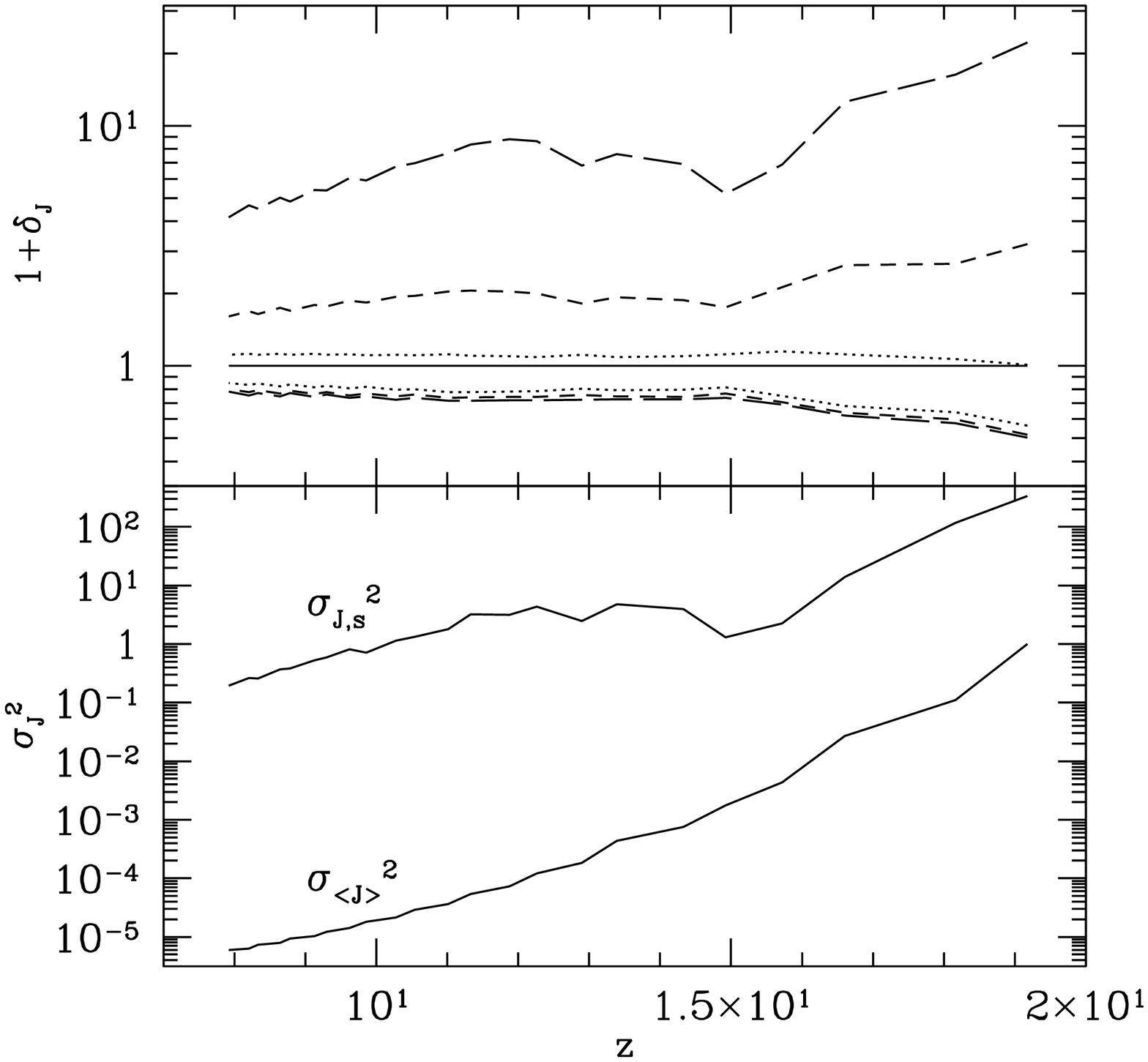}

\caption{({\em left}) Probability distribution function (``PDF'') of 
$J_{{\rm LW},\,21}$ for the $203^3$ grid cells inside a box of comoving 
size $35\, h^{-1}\,{\rm Mpc}$ at different redshifts. Numbers on 
individual curves represent the corresponding redshifts. ({\em right})
The top panel shows deviation of $J_{{\rm LW},\,21}$ from the 
globally-averaged $\langle J_{{\rm LW},\,21}\rangle$, expressed in terms 
of $1+\delta_{J}$. Around the curve of $\langle J_{{\rm LW},\,21}\rangle$
(solid), contours containing 68.27\% (dotted), 95.45\% (short dashed), 
and 99.73\% (long dashed) of the $J_{{\rm LW},\,21}$ distribution are 
shown. The sample variance of $J_{{\rm LW},\,21}$,
$\sigma_{J,\,s}^{2}$, is plotted together with the variance on the
average , $\sigma_{\langle J \rangle}^{2}$, in the bottom
panel. $\sigma_{\langle J \rangle}^{2}$ is dominated by the Poisson
error from the number of radiation sources rather than the
number of simulation grids, because the former is found to be much
smaller than the latter at all redshifts of our interest in our
simulation box. 
\label{fig:J21dev}} 
\end{figure*}

The PDF of $J_{{\rm LW},\,21}$ values on the simulation grid, plotted 
in Figure~\ref{fig:J21dev}, is also noteworthy. The volume-weighted
distribution of $J_{{\rm LW},\,21}$ is highly skewed. The deviation of 
$J_{{\rm LW},\,21}$ from 
the global average, $\langle J_{{\rm LW},\,21}\rangle$, is small when 
$J_{{\rm LW},\,21}<\langle J_{{\rm LW},\,21}\rangle$ (or $\delta_{J}<0$, 
where $\delta_{J}\equiv(J-\langle J\rangle)/\langle J\rangle$).
The minimum fluctuation is $\delta_{J,\,{\rm min}}\approx-0.5$ at our 
starting redshift, approaching zero as time goes on. Roughly 
speaking, $J_{{\rm LW},\,21}\simeq\langle J_{{\rm LW},\,21}\rangle$ 
when $\delta_{J}<0$. In contrast, strong deviations are observed in 
regions with $\delta_{J}>0$. Accordingly, most of the fluctuation 
involves $J_{{\rm LW},\,21}>\langle J_{{\rm LW},\,21}\rangle$. At 
all redshifts, $J_{{\rm LW},\,21}$ shows a variation of about two 
orders of magnitude from the minimum to the maximum, and about an 
order of magnitude variation at the $99.73\%$ level. 

The (sample) variance of $J_{{\rm LW},\,21}$, 
$\sigma_{J}^{2}\equiv\langle\delta_{J}^{2}\rangle$, also plotted in 
Figure~\ref{fig:J21dev}, does not evolve in a monotonic way. It 
starts out very large at $z\approx19.2$ and slowly decreases in 
time until $z\approx15$, then increases again until $z\approx12$ 
(except for a brief, sudden drop at $z\approx13$), and finally 
decreases again
after $z\approx12$. This limited range of redshift where $\sigma_J^2$ 
reverses its evolutionary trend, $z\approx[15-12]$, corresponds to 
the epoch when sources start to form inside massive halos with 
$M\ge10^{9}{\rm M}_{\odot}$, which are not subject to suppression and
``self-regulation'' (see Figure~\ref{fig:jnu}). We suspect that the 
small sudden drop of $\sigma_{J}^{2}$ at $z\approx13$, however, is just
due to the inherent cosmic variance. 

\begin{figure*}[ht]
\includegraphics[width=0.5\textwidth]{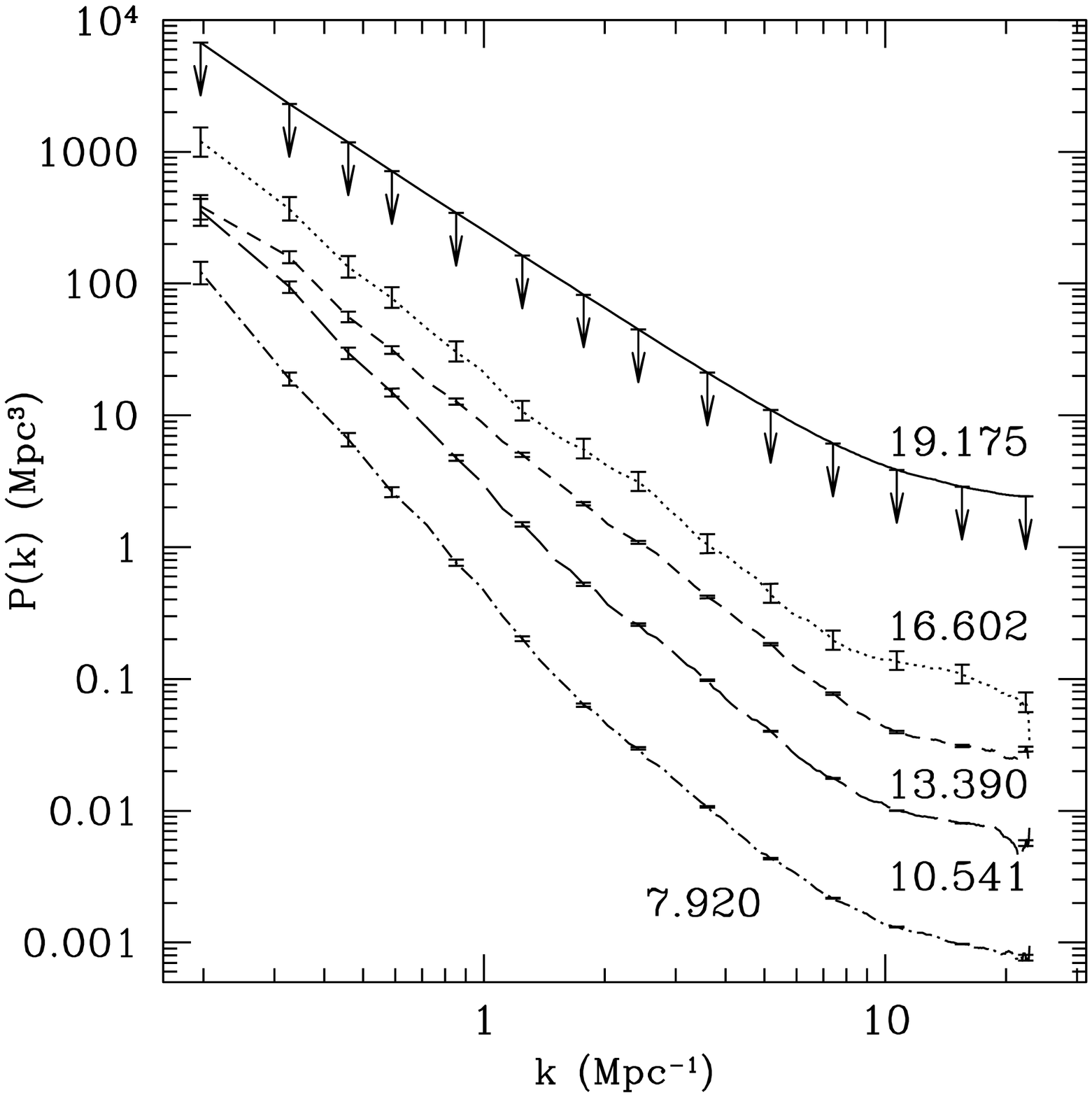}
\includegraphics[width=0.5\textwidth]{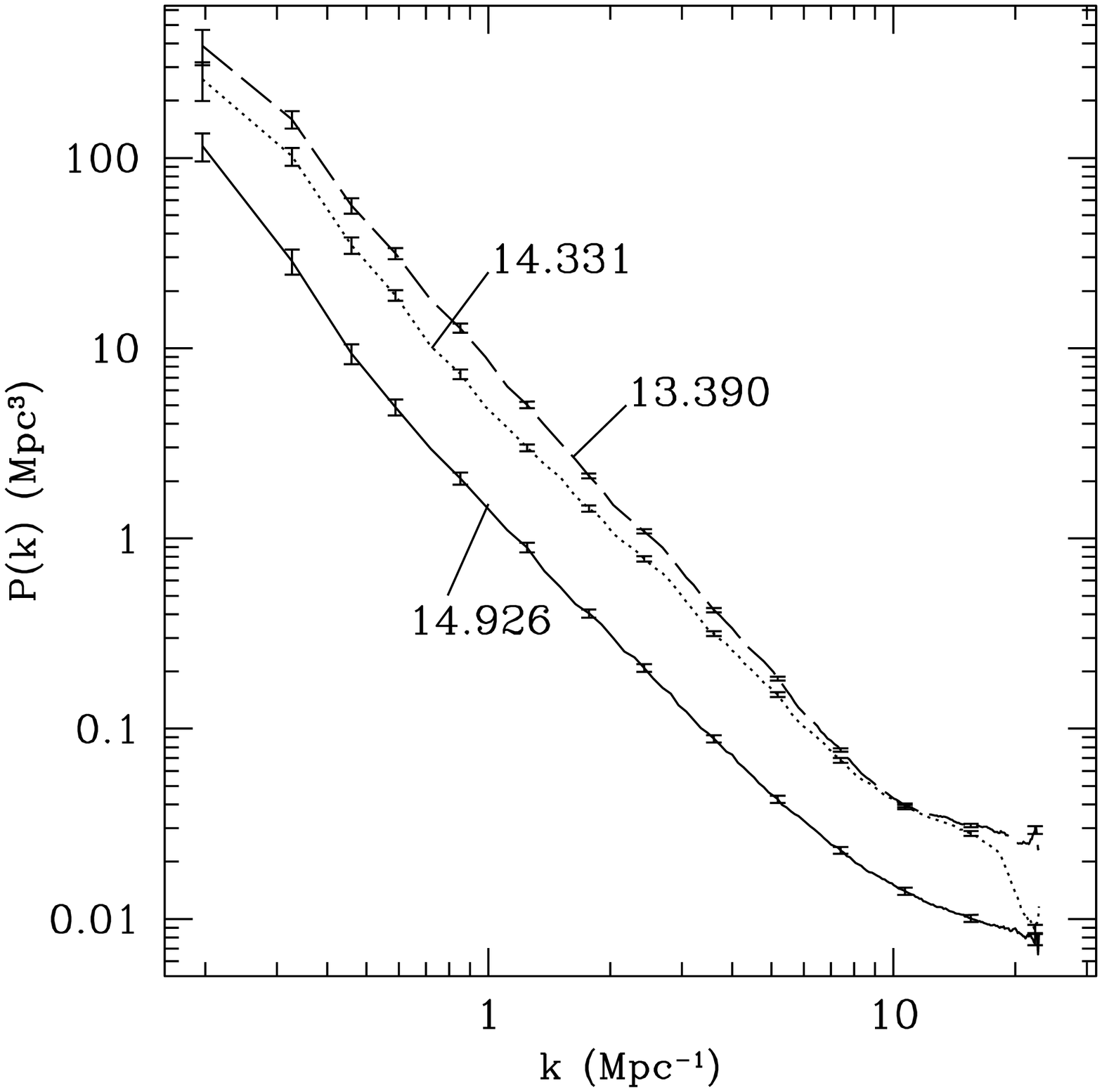}

\caption{({\em left}) Power spectrum $P(k)$ of LW background 
fluctuations $\delta_{J}$ at different redshifts. Numbers next 
to plotted curves represent corresponding redshifts. ({\em right}) 
$P(k)$ at limited range of redshifts, $z\approx15-13.4$. These 
show a reversed evolutionary track, increasing in time, compared
to $P(k)$'s plotted on the left panel, decreasing in time. Note that
this trend can also be seen in Figure~\ref{fig:J21dev}. Error bars on
both plots represent variance of $P(k)$ due to the finite number of
wavenumbers and the finite number of radiation sources. At $z=19.175$,
there is only one radiation source in the box, and the corresponding
power spectrum is roughly identical to the upper limit of $P(k)$. 
\label{fig:power}}
\end{figure*}

The power spectrum of $\delta_{J}$, $P(k)$, is shown in 
Figure~\ref{fig:power}. Large scales (small $k$) have more power than 
small scales (large $k$) do. The shape of $P(k)$ is rather complex, and 
a simple power-law provides a poor fit. Nevertheless, if a local 
power-law fitting is used, its power index becomes steeper in time. The 
overall amplitude of $P(k)$ decreases in time almost monotonically, 
except for an increase from $z\approx15$ to $z\approx12$, which is 
consistent with the trend seen in the evolution of $\sigma_{J}^{2}$ in 
Figure~\ref{fig:J21dev}. As for the normalization constant of $P(k)$, 
we follow the convention that the variance of $\delta_{J}$ is given by
\begin{equation}
\sigma_{J}^{2}\equiv\langle\delta_{J}^{2}\rangle
    =\frac{1}{2\pi^{2}}\int_{0}^{\infty}P(k)k^{2}dk,
\label{eq:sigma_power_cont}
\end{equation}
in the limit in which the size of the volume over which the average is 
calculated becomes infinitely large.

\begin{figure*}[ht]
\includegraphics[width=0.45\textwidth]{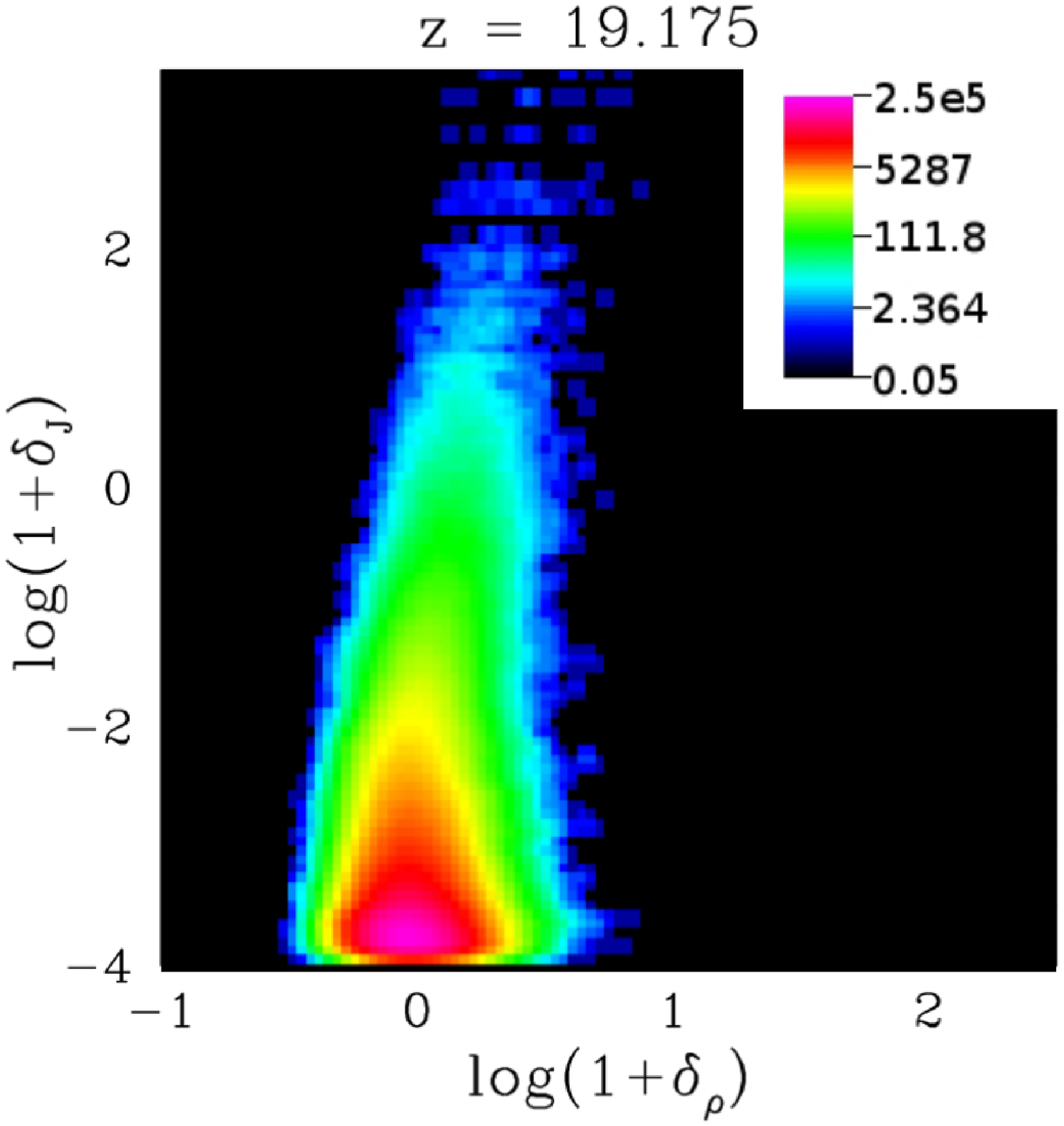}
\includegraphics[width=0.45\textwidth]{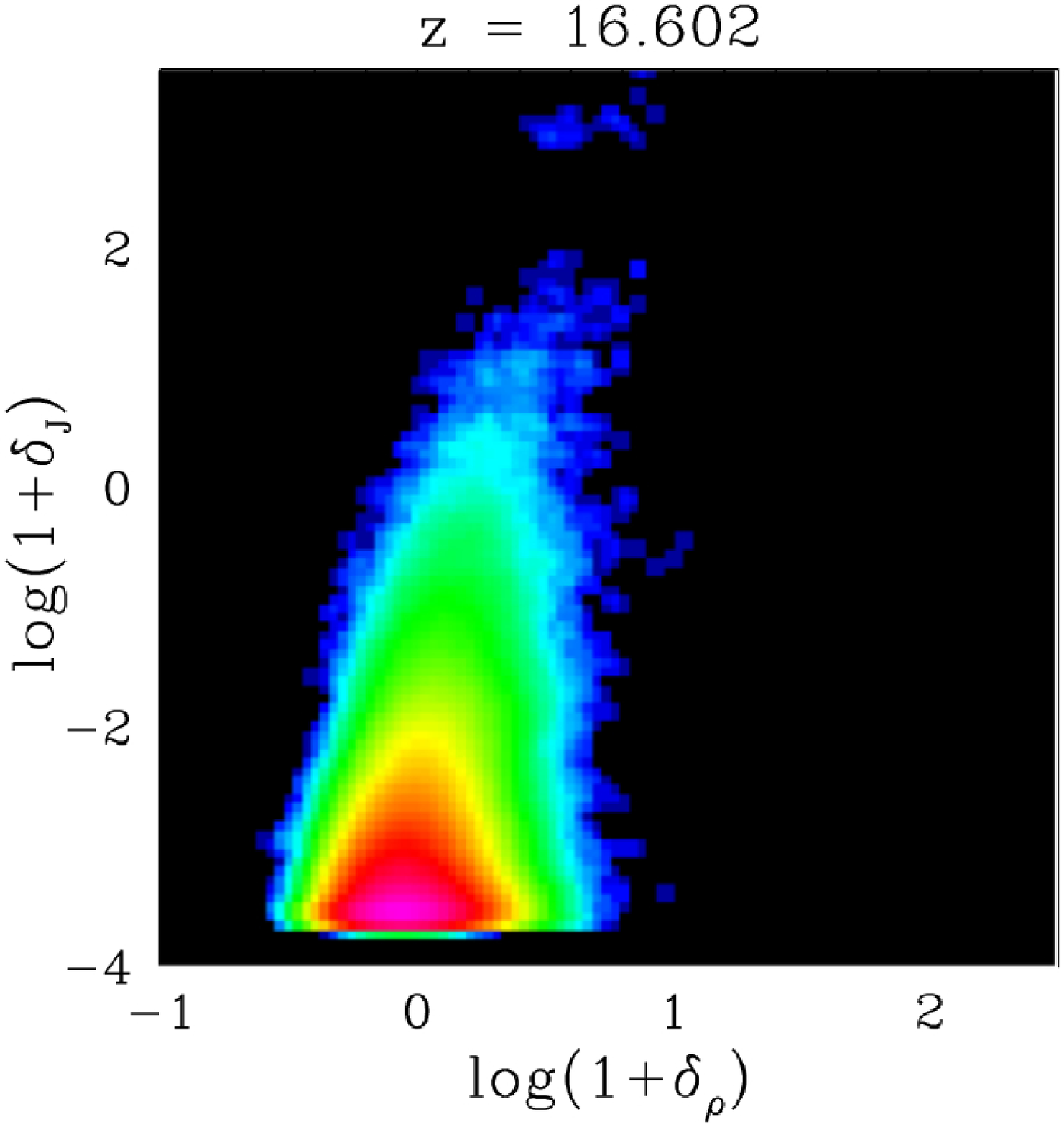}

\includegraphics[width=0.45\textwidth]{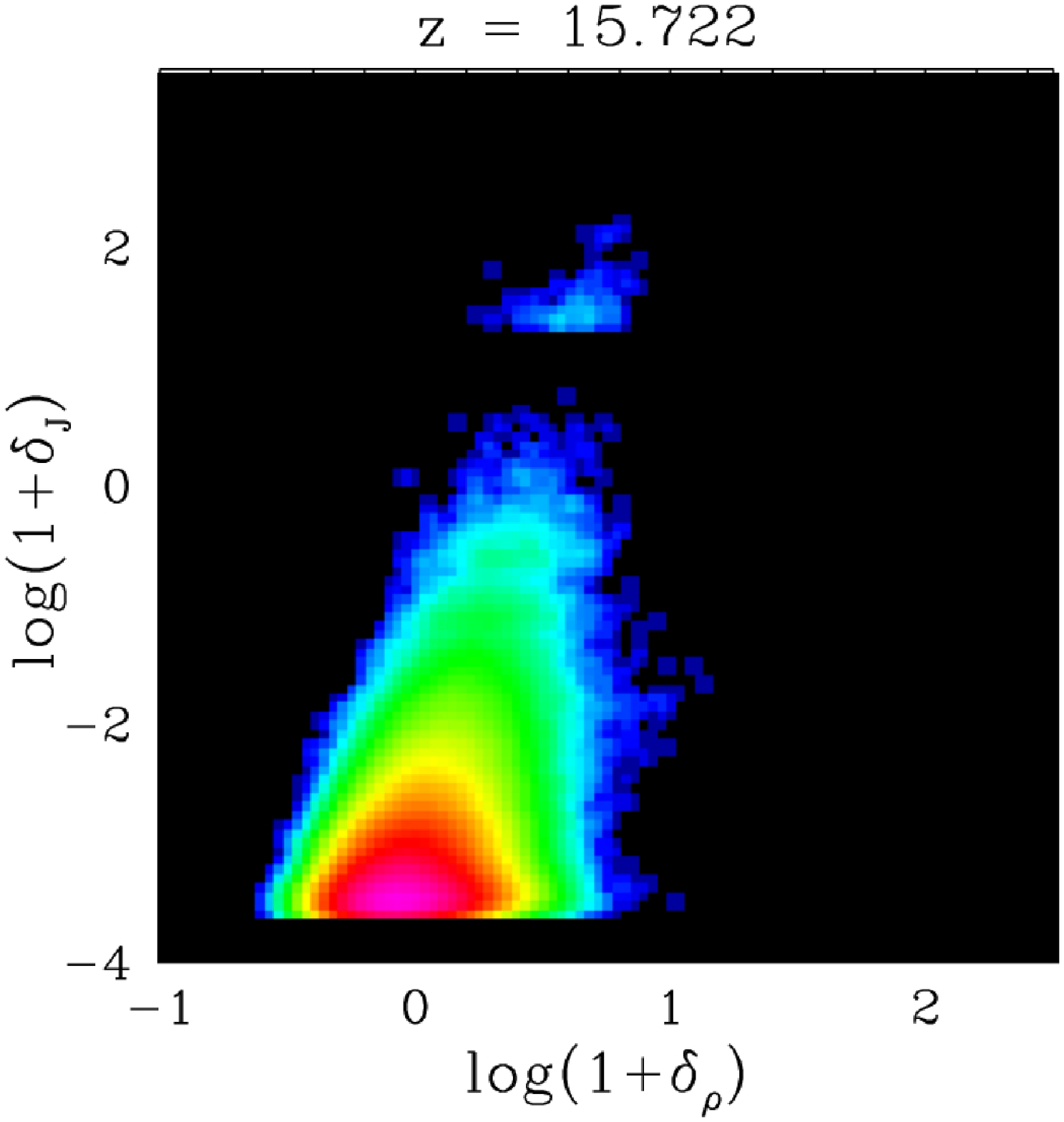}
\includegraphics[width=0.45\textwidth]{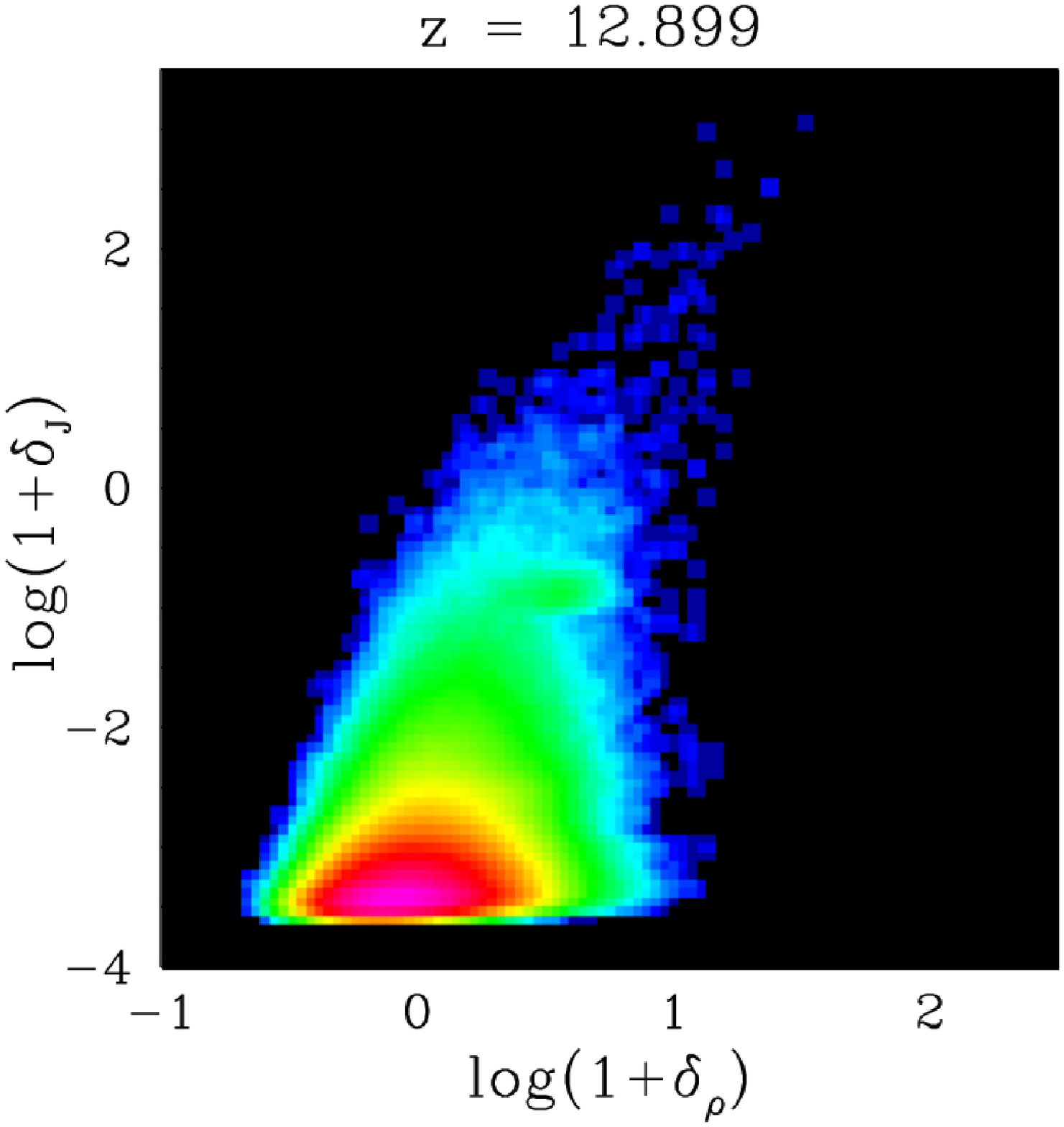}

\includegraphics[width=0.45\textwidth]{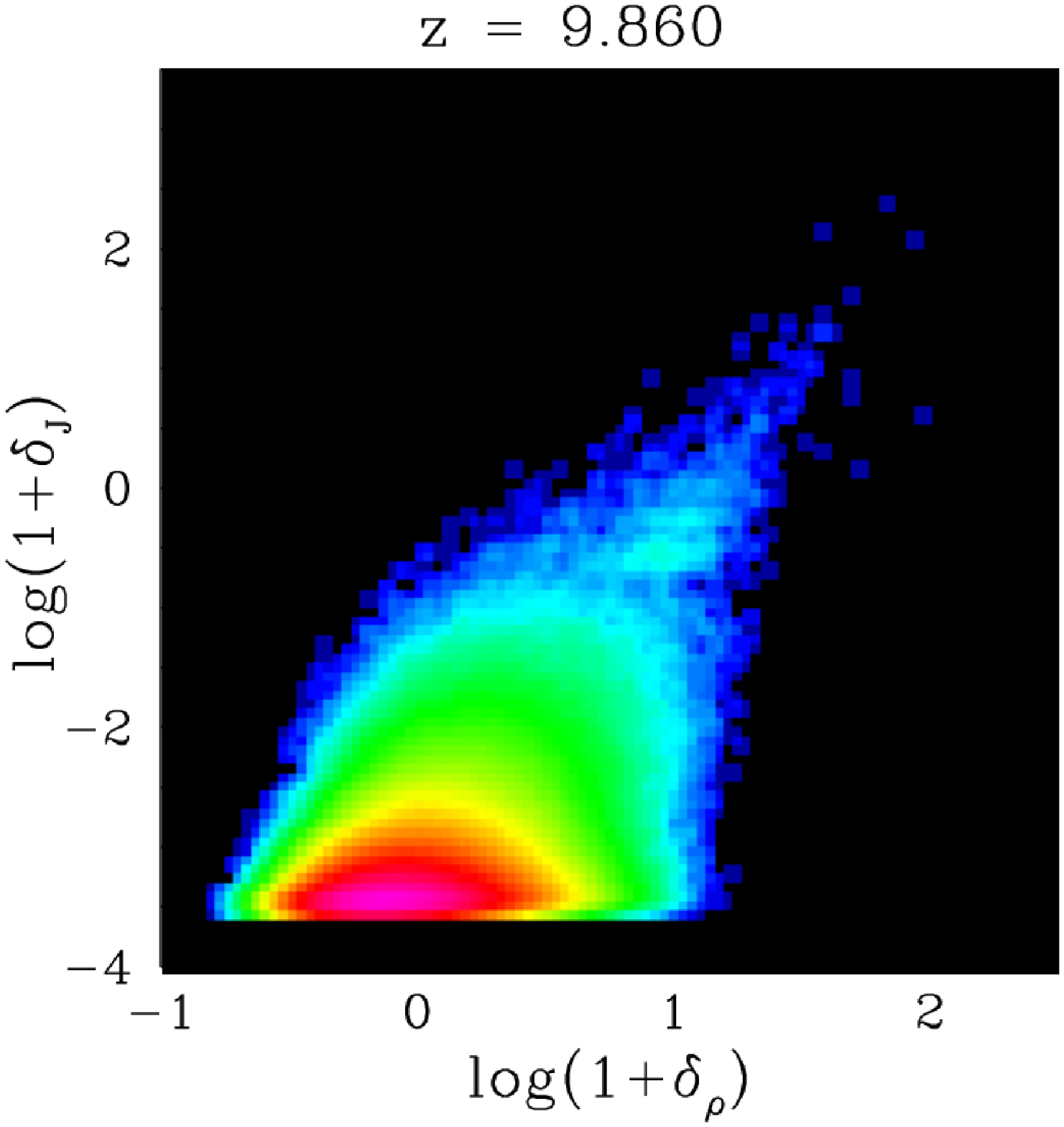}
\includegraphics[width=0.45\textwidth]{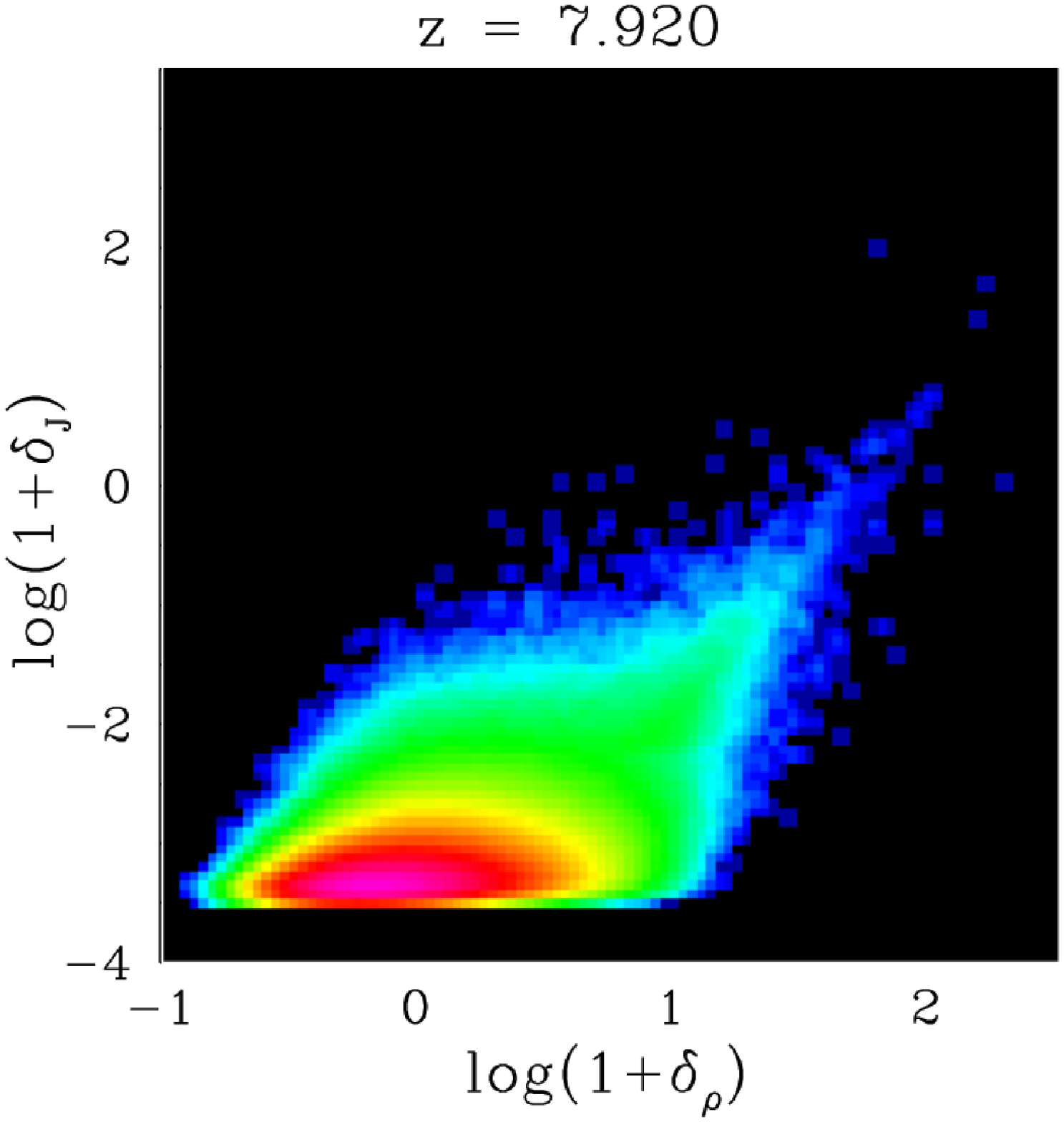}

\caption{Correlation of $J_{{\rm LW},\,21}$ and $\rho_{m}$ at different epochs,
  depicted by the number of cells at given $\delta_{J}$ and
$\delta_{\rho_{m}}$, with uniform bin-size of $\log (1+\delta_\rho)$
  and $\log (1+\delta_{J_\nu})$. Redshifts of these panels match those
  of Fig \ref{fig:J21cuts}. The inset in the top left panel shows the
  color scheme to depict the number of grid cells.
\label{fig:J21-rho-correlation}}
\end{figure*}

What causes such huge fluctuations within the simulation box, even when
its size is smaller than $r_{{\rm LW}}$? The answer is straightforward:
radiation sources cluster on scales smaller than $r_{{\rm LW}}$, and 
their spatial clustering generates fluctuations in the LW background 
that are not washed out even after the fluxes from all sources within
a distance $r_{\rm LW}$ are added up.
The original assumption that the LW background would evidence only 
small fluctuations misses this important ingredient. As simulated and 
noted by \citet{2006MNRAS.369.1625I} and others, patchiness of cosmic 
reionization, itself, strongly reflects the source-clustering effect. 
Sources cluster in high-density regions and will also produce a 
stronger LW background nearby, therefore. Such a correlation between the
LW fluctuations and the matter density fluctuations is depicted in 
Figure~\ref{fig:J21-rho-correlation}. 

Since reionization is ``inside-out'' according to these simulations (i.e.
the high-density regions ionize first), there is also a correlation between 
the H~II regions and the regions of higher-than-average LW intensity, as
seen in Figure~\ref{fig:J21cuts}. In fact, an animated sequence of maps 
like those in Figure~\ref{fig:J21cuts} shows that isocontours of 
$J_{\rm LW,21}$ start out centered on the same density peaks where H~II 
regions first appear. However, the isocontours of $J_{\rm LW,21}$ expand 
more rapidly than the ionization fronts (``I-fronts'') that define the H~II
boundaries, overtaking the I-fronts and expanding beyond the H~II regions.

\begin{figure}[ht]
\includegraphics[width=0.5\textwidth]{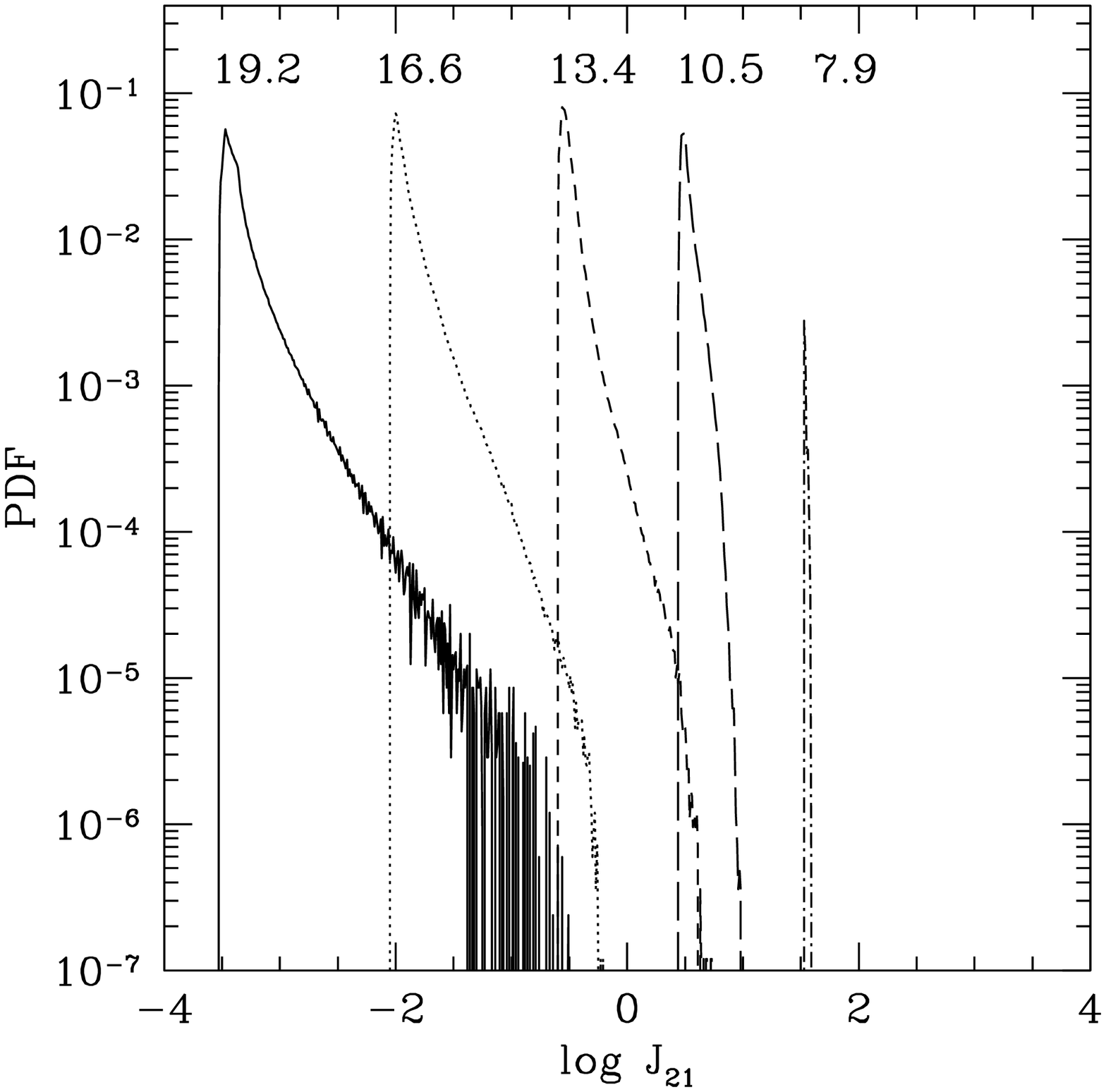}
\caption{Probability distribution of $J_{{\rm LW},\,21}$ in neutral cells
  only. Compare this plot to Fig \ref{fig:J21dev}. Decrease of overall
  area of PDF in time simply reflects the fact that the total volume of
  neutral regions decreases as cosmic reionization
  proceeds.\label{fig:histo_neu}}  
\end{figure}

Finally, we focus our attention on the LW intensity field in the neutral 
regions alone. Some fraction of the Pop III stars, including the very 
first ones, are believed to have formed inside cosmological minihalos. 
Minihalos in {\it ionized} regions are less likely to have formed stars,
however. If a minihalo formed in the ionized region, ``Jeans-mass filtering''
would have meant that it formed of dark matter only, devoid of baryons 
\citep[e.g.][]{1994ApJ...427...25S}. On the other hand, if a 
{\it pre-existing} minihalo found itself inside an ionized region
that had expanded to overtake it before the minihalo had yet formed 
a star, such a minihalo would have photoevaporated in the ionized region 
\citep{2004MNRAS.348..753S,2005MNRAS...361..405I}. Accordingly, star 
formation in minihalos in ionized regions was suppressed even more 
readily and further than in the low-mass atomic-cooling sources, 
halos with mass $M\geq10^8$. As far as star formation in minihalos is 
concerned, therefore, the {\it neutral} regions of the IGM are of particular 
interest. We find (see Figure~\ref{fig:histo_neu}) that early-on, the PDF 
distribution of $J_{{\rm LW},\,21}$ in the neutral regions is very similar 
to that overall, except with the highest-flux tail of the distribution cut 
off, since those are the regions in the immediate vicinity of the ionizing 
sources, which are ionized first.  The standard deviation of the PDF of
$J_{{\rm LW},\,21}$ decreases in time more rapidly for the neutral regions 
than for that overall. During the late stages of reionization, the values 
of $J_{{\rm LW},\,21}$ in these neutral cells converge to the average value, 
$\langle J_{{\rm LW},\,21}\rangle$, and the fluctuations largely disappear.

\section{Conclusion and Future Prospects}
\label{sec:conclusions}

We have, for the first time, calculated the inhomogeneous background of 
${\rm H}_{2}$ dissociating UV radiation caused by the same sources which 
reionized the Universe in a large-scale radiative transfer simulation of 
cosmic reionization. The UV continuum emitted below 13.6 eV by each 
source was transferred through the IGM, attenuated by atomic H Lyman 
series resonance lines, to predict the evolution of the inhomogeneous
radiation background in the energy range of the LW band of ${\rm H}_{2}$ 
between 11 and 13.6~eV. This required us to transfer the radiation from 
the many thousands of source galactic halos in our simulation, which 
resolved all halos of mass above $10^8M_\odot$ in a comoving volume of 
$(50\,\rm Mpc)^3$, from each source to each of the millions of grid cells 
of our reionization simulation. 
To accomplish this, we developed a novel method to calculate the attenuation 
of LW band photons from individual sources by H Lyman series resonance lines,
an otherwise prohibitively expensive, multi-frequency calculation, in a very
fast way, instead, without an explicit multi-frequency operation. This was 
achieved by a grey opacity approximation, the {}``picket-fence'' modulation 
factor, a simple function of the comoving distance between a source and an 
observer, which represents the frequency-averaged attenuation of LW band 
photons. We also explicitly accounted for the effect of the finite 
light-crossing time between sources and observers, by constructing the 
conformal space-time diagram of all reionization sources and observer grid
cells and finding the intersections between source world lines and 
observer past light cones.

Our results here demonstrate that the rise of the cosmic LW background 
to levels which have previously been identified as the threshold for 
dissociating $\rm H_2$ and thereby suppressing star formation inside 
minihalos occurs well before the epoch of reionization is very advanced.  
Not only is this the case for the mean LW background intensity, as 
anticipated by earlier estimates based upon a homogeneous universe 
approximation, but it is even more the case for the inhomogeneous 
background. The first regions to form halos are 
the regions with the highest mean overdensity, and we show that this 
is where the LW background rises the fastest and is at the highest 
levels.  This means that our
assumption here, for simplicity, that reionization is dominated
by the atomic cooling halos, and that minihalos are sterilized by the
rising LW background before they can contribute significantly to the
ionizing and, accordingly, the LW background, as well, is self-consistent. 

On the other hand, since there are also some minihalos that form far
from the density peaks around which the halos cluster, we might also
expect that there are {\em some} minihalos that form far from the peaks 
in the LW background, before their local LW intensity has risen to the
threshold level for $\rm H_2$ suppression.  In the future, it will be 
interesting to consider the possible role of these minihalos that 
form in places where the LW background is not high enough to suppress 
their star formation. 

Our result can be used for various applications. For example,
fluctuating ${\rm H}_2$ abundance in IGM and cosmological halos may be
calculated inside our simulation box. This would at least require
implementing reaction rates of neutral and ionic species of H, He, and
${\rm H}_2$. One may have to run many small-box simulations similar to those of
\citet{2003ApJ...592..645Y}, in order to calculate ${\rm H}_2$ abundance and
track source formation inside cosmological halos under fluctuating
$J_{\rm LW}$ calculated in this paper. Similarly, once X-ray emitting
sources are properly populated in the simulation box, a composite
effect of negative (due to photodissociation by UV) and positive (due
to partial ionization by X-ray) feedback may be studied as well.

There is the possibility that source formation inside the more massive,
atomic-cooling halos is affected by the LW background, too, because 
molecular cooling takes place inside these halos in the following way.
When the gas cools from the ionized state in these halos, it does so
out of equilibrium (e.g. \citealt{1987ApJ...318...32S};
\citealt{1992ApJ...386..432K}): gas cools faster than it recombines, 
so even after atomic H cooling has reached its typical saturated 
phase at $T\lesssim 10^4 \,{\rm K}$, there still remains a significant trace 
amount of electrons. Gas-phase reactions can then create ${\rm H}_2$ 
with the help of these electrons, which can further cool the gas down 
to $T\sim 200 \,{\rm K}$. Even though ${\rm H}_2$ at this stage can 
become self-shielded against the 
UV dissociating background (\citealt{2002ApJ...569..558O}), a strong
enough background may nevertheless dissociate ${\rm H}_2$ and suppress 
star formation to some extent (HAR). The threshold LW intensity for 
such suppression inside these atomic cooling halos, 
$(J_{\rm  LW})_{\rm threshold, atomic}$, 
is believed to be much larger than $(J_{\rm LW})_{\rm threshold}$ for 
minihalos (HAR). This might, therefore, somewhat affect the history 
of cosmic reionization near the end of reionization, when even
$(J_{\rm LW})_{\rm threshold, atomic}$ has been reached.

Emissivity by the first stars that contribute to the cosmic near-infrared 
background (NIRB) may also be affected by the fluctuating LW background
calculated here. Observations of a strong NIRB excess over the known 
foreground have been reported (e.g. \citealt{2005ApJ...626...31M}), and 
this has been interpreted as a possible signature of the first stars 
(c.f. \citealt{2006Natur.440.1002M} and references therein), although 
the true identity of the contributing sources is currently under debate 
(e.g. \citealt{2007ApJ...666L...1K}; \citealt{2007ApJ...666..658T}). Our 
work will impact the interpretation of both the homogeneous and inhomogeneous 
components of the NIRB, because predictions of the first star formation 
should be strongly affected by our results. 

We argued in \S~\ref{sec:Methodology} that we are justified in neglecting
${\rm H}_{2}$ self-shielding by the IGM in calculating the opacity 
to LW photons. This was justified by the fact that the concentration of 
$\rm H_2$ in the pre-reionization IGM was only $\sim10^{-6}$ and, as such, 
$\tau_{H_2}\lesssim1$. However, we can now use hindsight to see that our 
neglect of $\tau_{H_2}$ is even more self-consistent than that estimate 
would suggest. Hydrogen molecules in the IGM are quickly dissociated by 
the rising LW background. It is easily seen from Figures~\ref{fig:nn}, 
\ref{fig:J21cuts} and \ref{fig:J21dev} that 
$J_{{\rm LW},\,{21}} \gtrsim 10^{-2}$ at $z\simeq 20$ in the vicinity 
of atomic cooling halos, and the mean value quickly rises to 
$\langle J_{{\rm LW},\,{21}}\rangle \simeq 10^{-2}$ and above. When
the mean IGM prior to 
reionization (with temperature $T_{\rm IGM}\simeq 369(1+z)^{2}/135^{2}$, 
hydrogen number density $n_{\rm H}\simeq 1.7 \times 10^{-3}\,(1+z)^{3}
/ 21^{2}\,{\rm cm}^{-3}$, and  
$y_{{\rm H}_2}\simeq 2\times 10^{-6}$ when there is no LW background)
is exposed to the LW intensity $J_{{\rm LW},\,{21}} = 10^{-2}$ from 
$z=20$ onward, for example, its molecular fraction drops 
to $y_{{\rm H}_2}\simeq 2\times 10^{-7}$ by $z\simeq 17$. Even if 
${\rm H}_{2}$ self-shielding were marginally important with 
$y_{{\rm H}_2}\simeq 2\times 10^{-6}$ at high redshift
(\citealt{2002ApJ...575...49R}), therefore, this will quickly become 
negligible when molecules are dissociated to the level $y_{{\rm H}_2}\simeq
2\times 10^{-7}$, which would occur at $z\simeq 17$ from the LW 
background contributed by the atomic-cooling halos alone. 
Peculiar motion of gas elements will weaken self-shielding
even further (e.g. \citealt{2001ApJ...548..509M}). 
\citet{2007ApJ...665...85J}, on the other hand, claim that relic H~II 
regions created by
the first stars can recombine and generate abundant ${\rm H}_{2}$ before
being exposed to other external sources, through the nonequilibrium
$\rm H_2$-formation mechanism described above (\citealt{1987ApJ...318...32S};
\citealt{1992ApJ...386..432K}). \citet{2007ApJ...665...85J} find,
however, that this boost of ${\rm H}_2$ abundance is delayed 
significantly when $J_{{\rm LW},\,{21}} \gtrsim 10^{-2}$, which is easily 
satisfied in the mean IGM at $z\simeq 17$ and in the vicinity of atomic 
cooling halos at $z\simeq 20$.
This suggests that the proposed nonequilibrium enhancement of the $\rm H_2$ 
concentration in the IGM inside relic H~II regions is not likely to 
provide a significant enough $\rm H_2$ opacity to shield the IGM through
most of the EOR.

\ack

This study was supported in part by KICOS through the grant
K20702020016-07E0200-01610 provided by MOST, 
NSF grant AST
0708176, NASA grants NNX07AH09G and  
NNG04G177G, Chandra grant SAO TM8-9009X, Swiss National Science 
Foundation grant 200021-116696/1, and Swedish Research Council grant 
60336701. The authors acknowledge the Texas Advanced Computing Center 
(TACC) at The University of Texas at Austin for providing HPC
resources that have contributed to the research
results reported in this paper.


\end{document}